\documentclass[aps,pra,twocolumn,letterpaper,superscriptaddress, floatfix]{revtex4} 

\usepackage{dsfont}
\usepackage{amsmath}
\usepackage{amsfonts}
\usepackage{amssymb}
\usepackage{bbm}
\usepackage{bm}
\usepackage{graphicx}
\usepackage{multirow}
\usepackage[latin1, utf8]{inputenc}
\usepackage{color}
\usepackage{xcolor}
\usepackage{ulem}

\usepackage{listings}

\usepackage[colorlinks=true,citecolor=black,urlcolor=black,linkcolor=black]{hyperref}

\usepackage[caption=false, labelformat=empty]{subfig}

\newcommand{\C}{\mathbbm{C}}
\newcommand{\R}{\mathbbm{R}}

\renewcommand\Re{\operatorname{Re}}

\newcommand{\br}{{\bm r}}

\newcommand{\blambda}{{\bm\lambda}}
\newcommand{\BLambda}{{\mathbf\Lambda}}

\makeindex
\begin{document}

\title{Optimal control of Bose-Einstein condensates in three dimensions} 

\author{J.-F.~Mennemann}
\email{mennemann@acin.tuwien.ac.at}
\affiliation{Automation and Control Institute, Complex Dynamical Systems Group, Vienna University of Technology, 
Gusshausstrasse 27-29, 1040 Vienna, Austria}
\affiliation{Department of Mathematics, University of Vienna, 1090 Vienna, Austria}


\author{D.~Matthes}
\affiliation{Zentrum Mathematik, Technische Universit\"at M\"unchen, D-85747 Garching, Germany}

\author{R.-M.~Weish\"aupl}
\affiliation{Department of Mathematics, University of Vienna, 1090 Vienna, Austria}

\author{T.~Langen} 
\email{tim.langen@colorado.edu}
\affiliation{Vienna Center for Quantum Science and Technology, Atominstitut, TU Wien, Stadionallee 2, 1020 Vienna, Austria}
\affiliation{JILA, University of Colorado and NIST, Boulder, Colorado 80309, USA}

\begin{abstract}
Ultracold gases promise many applications in quantum metrology, simulation and computation. In this context, optimal control theory (OCT) provides a versatile framework for the efficient preparation of complex quantum states. However, due to the high computational cost, OCT of ultracold gases has so far mostly been applied to one-dimensional (1D) problems. Here, we realize computationally efficient OCT of the Gross-Pitaevskii equation (GPE) to manipulate Bose-Einstein condensates in all three spatial dimensions. We study various realistic experimental applications where 1D simulations can only be applied approximately or not at all. Moreover, we provide a stringent mathematical footing for our scheme and carefully study the creation of elementary excitations and their minimization using multiple control parameters. The results are directly applicable to recent experiments and might thus be of immediate use in the ongoing effort to employ the properties of the quantum world for technological applications.
\end{abstract}

\date{\today}

\maketitle

\section{Introduction}

Over the last decade, the ever increasing experimental toolbox of atomic, optical and molecular physics has lead to an exciting improvement in the control and understanding of complex quantum systems~\cite{Bloch08}. 
Recently, this has resulted in an important shift of paradigm. While quantum systems were previously mostly studied to check the validity of theoretical models, interest has now increased in their manipulation for specific technological applications. 
Prototypical examples for this shift of paradigm are atomic interferometers for quantum enhanced metrology~\cite{Gross2010,Lucke2011,Riedel2010}, atomic field probes~\cite{Ockeloen2013} and microscopes~\cite{Wildermuth05,Aigner08}, inertial sensors~\cite{Geiger2011}, atomic clocks~\cite{Bloom14}, or applications in quantum computing~\cite{Calarco2000,Kielpinski2002} and quantum simulation~\cite{Bloch12}. 

In many cases, these applications rely on the controlled preparation of a well-defined quantum many-body state with particular properties. 
One of the key experimental challenges is thus the efficient transfer of a system to such a state. 
Optimal control theory (OCT) is a mathematical tool to devise control strategies for this transfer~\cite{Peirce1988}. 
It is well studied in many physical systems, ranging from atoms and molecules to solid-state systems~\cite{Koch2004,Rabitz2000,Borzi2002,Hohenester2006,Nobauer2014}. 

In this work, we apply it to the control of a dilute atomic Bose-Einstein condensate (BEC), a system which is well described by the three-dimensional (3D) Gross-Pitaevskii equation (GPE)~\cite{Pethick2002,Dalfovo99}. Such BECs form a versatile experimental platform for the storage, manipulation and probing of interacting quantum fields with high precision~\cite{Bloch08}. 
In a seminal work Hohenester et al.~\cite{HoReBoSc07} demonstrated that OCT~\cite{Peirce88} provides a highly efficient way to realize the transfer of a BEC to a target state, vastly outperforming more simple schemes. 
In this context, it has also been shown that OCT is robust against fluctuations and decoherence, and can also specifically take into account experimental constraints~\cite{Jager2013}. This has recently lead to first experimental demonstrations~\cite{Bucker2011,vanFrank2014}.  

OCT of BECs has so far been mostly used in one-dimensional (1D) settings, as the computational cost scales exponentially with the number of dimensions~\cite{Ho2014}. 
However, many experimental situations can only approximately be described by a 1D model, potentially limiting the applicability of OCT to real-life situations. 
In the following we demonstrate the first OCT of a BEC in all three spatial dimensions.
We go beyond situations where a 1D approximation is feasible, thus significantly expanding the range of applicability of OCT for BECs. Moreover, we perform an analysis of the collective excitations that are created as a result of the control. These excitations are directly connected to the non-linear nature of the GPE and can only be fully captured and minimized in a 3D treatment including multiple control parameters. They are thus highly relevant in realistic experimental situations.

\section{The control problem}
We start with a brief review of OCT, as well as of the description of BECs in terms of the GPE.

\subsection{Gross-Pitaevskii equation}

The mean-field dynamics of a BEC is described by the GPE
\begin{equation}
\label{eq:gpe}
i \hbar \partial_t \psi = -\frac{\hbar^2}{2 m} \Delta \psi + V_\blambda \psi
+ g |\psi|^2 \psi
\end{equation}
where $\psi\equiv\psi(\br,t)$ denotes a complex-valued wave function, with initial condition
$\psi(\bm{r},0) \equiv \psi_0 \in L^2(\R^3;\C)$. Here, $V_\blambda\equiv V(\br,\blambda(t))$ is an external potential that is characterized by a single or several control parameters denoted by the vector $\blambda$. Assuming that the wave function is normalized to unity, the
coupling constant
$
g = N 4 \pi \hbar^2 a_s/m
$
is defined by the mass $m$, the s-wave scattering length $a_s$ and the number $N$ of atoms in the BEC. For example, for ultracold gases of ${}^{87}$Rb atoms, the atomic mass is given by $m=1.44\times10^{-25}\,$kg and the s-wave scattering length by $a_s = 5.24$\,nm. 
Measuring length in units of $l_0 = 1\times \mu$m, mass in units of the atomic mass 
and time in units of $t_0 = m l_0^2 / \hbar$, equation \eqref{eq:gpe} can be written as 
\begin{equation}
\label{eq:gpe2}
i \partial_t \psi = -\frac{1}{2} \Delta \psi + V_\blambda \psi
+ g |\psi|^2 \psi
\end{equation}
which is the starting point for our considerations below.

\subsection{Optimal control problem}
\label{subsec:optimal_control_problem}

We seek to find an optimal time-evolution of the $m$-component control parameter
\[
\blambda:(0,T)\to\R^m,\quad \blambda(0)=\blambda_0,\quad \blambda(T)=\blambda_T,
\]
which steers the system from the initial state $\psi_0$ at time zero to a desired state $\psi_d$ at final time $T$.
Without loss of generality we assume that $\psi_0$ and $\psi_d$ are ground state solutions of the stationary GPE
corresponding to the smooth external potentials $V_{\blambda_0}$ and $V_{\blambda_T}$ at times $t=0$ and $t=T$, respectively,
with fixed parameters $\blambda_0$, $\blambda_T$.
To find the time evolution we apply well-known techniques from optimal control theory~\cite{HoReBoSc07}. 
As cost functional, we use
\begin{equation}
  \label{eq:cost_functional}
  J(\blambda,\psi) = \frac12\left(1-|\langle\psi_d,\psi(T)\rangle|^2\right) + \frac\gamma2\int_0^T |\partial_t\blambda(t)|^2\, dt,
\end{equation}
where $\langle u, v \rangle = \int_{\R^3} u(\bm{r})^* v(\bm{r}) \,d\bm{r}$ denotes 
the standard scalar product of $u,v \in L^2(\mathbbm{R}^3; \C)$.
The definition \eqref{eq:cost_functional} is the generalization of the functional 
used in Refs. \cite{HoReBoSc07, WiBo08, BueBeFr13, Jager2013, JaeHo2013, JaeReGoKoHo2014, Ho2014}
to a multi-component control parameter $\blambda$.
The first term in $J$ measures the proximity of $\psi$ to the desired state $\psi_d$ at the end of the steering process.
The expression $\widetilde{\mathcal F}(\psi)=1-|\langle\psi_d,\psi\rangle|^2$ is known as the infidelity and provides a measure for the difference of $\psi$ and $\psi_d$. In detail, it quantifies the $L^2$-norm of $\psi$'s component that is orthogonal to $\psi_d$. The second term regularizes the control trajectory to account for the fact that parameters can never be changed infinitely fast in a real experiment. Here, $\gamma > 0$ sets the penalty for fast variations of $\blambda(t)$. For our examples below we find that already a very small value $\gamma = 1 \times 10^{-6}$ yields a satisfactory regularization. 

Our goal is to minimize $J(\blambda, \psi)$ subject to the constraint that 
$\psi$ solves the GPE (Eq.~\eqref{eq:gpe2}) with the initial condition given by the respective $\psi_0$.
To this end, one introduces the Lagrange function
\begin{align}
\label{eq:Lagrange_function}
  L(&\blambda,\psi,p)
  = 
  J(\blambda, \psi)+ \\
  &\Re \int_0^T \int_{\R^3} p^*\big(i \partial_t \psi 
  + \frac12\Delta \psi - V_\blambda \psi - g |\psi|^2 \psi \big) \,dt\,d\bm{r}\nonumber
\end{align}
where $p(\bm{r},t)$ acts as a generalized Lagrange multiplier~\cite{Peirce88}.
At a local minimum $(\blambda,\psi,p)$ of $J$, 
all three variational derivatives
$D_p L(\blambda, \psi, p) [\delta p]$, $D_\psi L(\blambda, \psi, p) [\delta \psi]$
and
$D_\blambda L(\blambda, \psi, p) [\delta \blambda]$
vanish for all admissible variations $\delta p$, $\delta \psi$ and $\delta \blambda$, respectively.
The corresponding three conditions constitute the optimality system
\begin{subequations}
  \label{eq:optimality_system}
  \begin{align}
    \label{eq:state_equation}
    i \partial_t \psi &= -\frac{1}{2} \Delta \psi + V_\blambda \psi + g |\psi|^2 \psi,
    \\
    \label{eq:adjoint_equation}
    i \partial_t p &= - \frac{1}{2} \Delta p + V_\blambda p + 2 g |\psi|^2 p + g \psi^2 p^*,
    \\
    \label{eq:control_equation}
    \gamma \frac{d^2}{dt^2} \blambda 
    &= -\Re \langle \psi, (\partial_\blambda V_\blambda) \, p \rangle,
  \end{align}
\end{subequations}
together with the initial and terminal conditions
\begin{subequations}
  \label{eq:initial_and_terminal_conditions}
  \begin{align}
    \label{eq:initial_condition_state_equation}
    \psi(0) &= \psi_0,
    \\
    \label{eq:terminal_condition_adjoint_equation}
    i p(T) &= -\langle \psi_d, \psi(T) \rangle \psi_d,
    \\
    \label{eq:boundary_conditions_lambda}
    \blambda(0) &= \blambda_0,
    \quad \blambda(T) = \blambda_T.
  \end{align}
\end{subequations}

\noindent
In general, no analytical solutions are available for \eqref{eq:optimality_system} with~\eqref{eq:initial_and_terminal_conditions}.
Here we use an iterative method to find a numerical approximation of the solution.
For this purpose it is useful to introduce the reduced cost functional
\begin{equation}
  \label{eq:reduced_cost_functional}
  \hat J(\blambda) = J(\blambda,\psi_\blambda),
\end{equation}
where $\psi_\blambda$ denotes the unique solution of the Gross-Pitaevskii equation for a given
control parameter curve $\blambda$.
The goal is to find a local (or, preferably, even global) minimizer $\blambda^*$ of $\hat J$.

The most straight-forward iterative procedure that can be employed is the method of steepest descent, 
\begin{equation}
  \label{eq:steepest_descent}
  \blambda^{k+1} = \blambda^k - \alpha^k \nabla \hat{J}(\blambda^k), \quad k=0,1,2,... . 
\end{equation}
To determine an appropriate step size $\alpha^k$, we perform a line search in each iteration:
\begin{equation}
\alpha^k = \operatorname*{arg\,min}_{\alpha} \,\hat{J}(\blambda^k - \alpha \nabla \hat{J}(\blambda^k)).
\end{equation}
Here the upper index denotes the iteration step. 
A comment is due on the use of the gradient $\nabla \hat{J}(\blambda^k)$ in~\eqref{eq:steepest_descent}.
Recall that the gradient of $\hat J$ at $\blambda$ with respect to a specific inner product $(\cdot,\cdot)_X$ 
on the space $X$ of admissible variations $\delta\blambda$
is the uniquely determined element $\nabla \hat{J}\in X$ such that $(\nabla \hat{J},\delta\blambda)_X = D_\blambda\hat J(\blambda)[\delta\blambda]$
for all admissible variations $\delta\blambda\in X$. The gradient thus depends sensitively on the choice of the inner product $(\cdot,\cdot)_X$ on $X$.
It has been pointed out already in Ref.~\cite{WiBo08} that any admissible variation $\delta\blambda$ 
must have a finite value in the penalty term, i.e., its weak time derivative $\partial_t\delta\blambda$ must be square-integrable on $(0,T)$,
and must respect the boundary conditions in~\eqref{eq:boundary_conditions_lambda}, i.e., $\delta\blambda(0)=\delta\blambda(T)=0$.
A natural choice for $(\cdot,\cdot)_X$ is thus the $H_0^1(0,T,\R^m)$-scalar product,
\begin{align}
  \label{eq:inner_product_H_0_1_R_m}
  (\bm{u},\bm{v}) := \int_0^T \partial_t \bm{u}(t) \cdot \partial_t \bm{v}(t)\,dt.
\end{align}
A calculation, which we present in the appendix, shows that this choice of $(\cdot,\cdot)_X$ yields
\begin{subequations}
	\label{eq:gradient}
 	\begin{align}
  	\label{eq:gradient_H_0_1}
  	\frac{d^2}{dt^2} \big[\nabla \hat{J}(\blambda)\big] 
 	& = 
  	\gamma \ddot{\blambda} + \Re \langle \psi, (\partial_\blambda V_\blambda) p  \rangle, \\
  	\big[\nabla \hat{J}(\blambda)\big](0)& =  \bm{0}, \\
  	\big[\nabla \hat{J}(\blambda)\big](T)& = \bm{0},
   	\end{align}
\end{subequations}
wherein $\psi$ and $p$ are solutions of~\eqref{eq:state_equation} and~\eqref{eq:initial_condition_state_equation}
or~\eqref{eq:adjoint_equation} and~\eqref{eq:terminal_condition_adjoint_equation}, respectively.
By definition, $\nabla\hat J$ vanishes at the boundaries $t=0$ and $t=T$,
and so the iteration~\eqref{eq:steepest_descent} preserves the boundary conditions~\eqref{eq:boundary_conditions_lambda}.
We emphasize that the seemingly canonical choice of $(\cdot,\cdot)_X$ as the standard $L^2$-scalar product
would not allow to specify boundary data for $\nabla\hat J$, which would result in
a severe loss of stability of the optimization algorithm.

\subsection{Implementation}

In the situations considered below we found that the 
method of steepest descent (see Eq. \eqref{eq:steepest_descent})
works reliably. 
However, using more advanced methods the number of iterations needed to ensure convergence of the algorithm can be reduced significantly. 
In fact, our solver is based on the non-linear conjugate gradient scheme of Hager and Zhang \cite{HaZh2005}, which has also been employed in Ref.~\cite{WiBo08} for optimal control of the one-dimensional GPE. We stress that all inner products and norms related to the non-linear conjugate gradient scheme need to be expressed in terms of the inner product given in 
Eq.~\eqref{eq:inner_product_H_0_1_R_m}. 

The reduced cost functional~\eqref{eq:reduced_cost_functional} needs to be evaluated several times per iteration.
Moreover, at the beginning of each iteration a gradient vector needs to be determined using Eq.~\eqref{eq:gradient}.
Solutions to the time-dependent GPE \eqref{eq:state_equation} and the 
adjoint equation \eqref{eq:adjoint_equation} are obtained via the time-splitting spectral method \cite{BaJaMa2003}.
Initial and desired final states for a given potential are found by imaginary time propagation. 

In order to accelerate the solving of the optimal control problem we perform all
computations on the graphics processing unit (GPU) of a powerful graphics card.
To speed up the calculations and ensure convergence of the algorithm we start each optimization with a coarse spatial grid and a relatively big time step $\Delta t$. 
The result for $\blambda$ is used as an input for another round of optimization on a finer grid. 
This procedure is repeated until the algorithm converges to a final time-evolution for $\blambda$. 
A detailed description of our implementation is given in Appendix B.

\section{Examples}
In the following we demonstrate the results of our scheme by considering three applications of increasing complexity, which are directly connected to recent experiments.

\subsection{Harmonic oscillator potential}
\label{sec:harmonic}

\begin{figure*}[htb]
\centering
\includegraphics[width=0.925\textwidth]{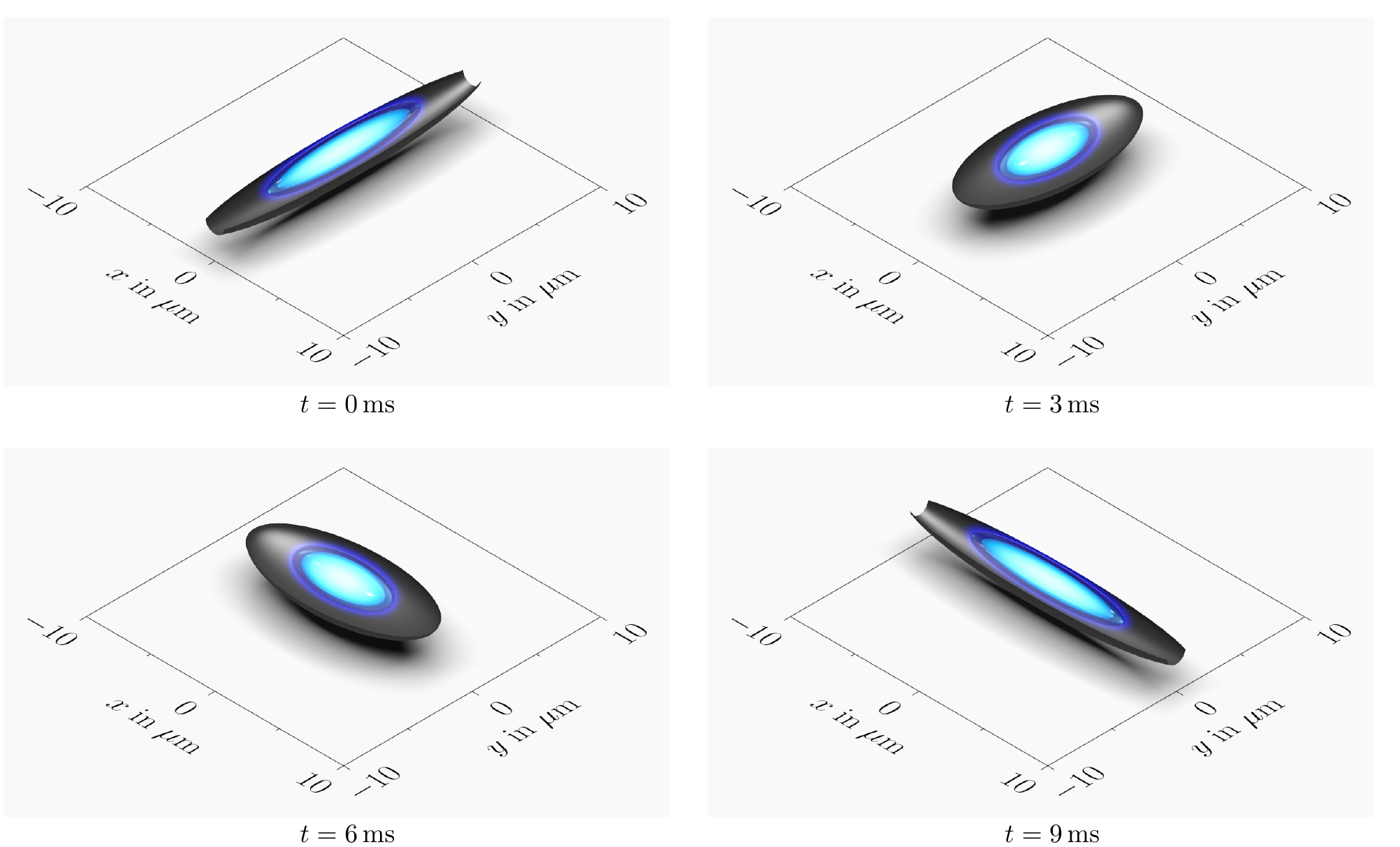}\\
\caption{Two-parameter optimal control of an elongated harmonic potential. The timescale of the control is $T=9$\,ms. Initially aligned along the $y$-direction, the condensate is dynamically transformed to be aligned along the $x$-direction. The black isosurface corresponds to the external trapping potential that is controlled using OCT, the blue isosurfaces visualize the atomic density. Note that for clarity only the lower half of the potential is shown. Also, here and throughout this work any trivial potential offset has been removed for simplicity and easier visualization. Its only effect is an overall phase shift of the wave function which is of no relevance to the optimization procedure. Animations of the full dynamics are available online \cite{videos}. 
}
\label{fig:harmonic_snapshots_3d}
\end{figure*}

In the first application we study a Bose-Einstein condensate in an elongated harmonic potential. Initially, the trap frequencies are chosen such that the condensate is aligned along the $y$ direction. 
Using a suitable time-evolution of the trap frequencies, we aim to rotate the condensate by $\pi/2$, while keeping it in the ground state of the external potential. 

An example of the transition is visualized in Fig.~\ref{fig:harmonic_snapshots_3d}. It can be understood as a toy example of a broad class of experimental protocols in which the trapping geometry is changed, e.g. to mode match different traps~\cite{Ketterle1999}, to (de)compress a trap~\citep{Schaff2011} or to transfer condensates into dynamical potentials for atomtronics~\cite{Seaman2007,Henderson2009}. Conceptually similar pulsed manipulations are also performed to focus BECs in time-of-flight expansion~\cite{Shvarchuck2002}.

\subsubsection{Trapping potential}

The harmonic potential in this example is given by
\[
V_\blambda (x,y,z) 
= 
\frac{m}{2} 
\Big( 
\left[\omega_x(\lambda_1)\right]^2 x^2 + \left[\omega_y(\lambda_2)\right]^2 y^2 + \omega_z^2 z^2
\Big),
\]
wherein the frequencies $\omega_x$ and $\omega_y$ can be set independently via the control parameters $\lambda_1$
and $\lambda_2$.
More precisely, 
we transform the external potential
from an initial configuration with $\omega_x = \omega_x^i$ and $\omega_y = \omega_y^i$ at time $t=0$
to a final configuration with $\omega_x = \omega_x^f$ and $\omega_y = \omega_y^f$ at the final time $t=T$.
To this end, we parametrize $\omega_x$ and $\omega_y$ as
\begin{align*}
\omega_{x}(\lambda_1) &= \omega_x^i + \lambda_1 (\omega_x^f - \omega_x^i),\\
\omega_{y}(\lambda_2) &= \omega_y^i + \lambda_2 (\omega_y^f - \omega_y^i),
\end{align*}
with
\begin{align*}
\lambda_1(0) &= 0, \quad \lambda_1(T) = 1,\\
\lambda_2(0) &= 0, \quad \lambda_2(T) = 1.
\end{align*}
We note that these parametrizations, as all others discussed below, are chosen as an example and can easily be adjusted to the parameters accessible in a specific experimental realization.  

\begin{figure*}[htb]
	\centering
	\includegraphics[width=0.965\textwidth]{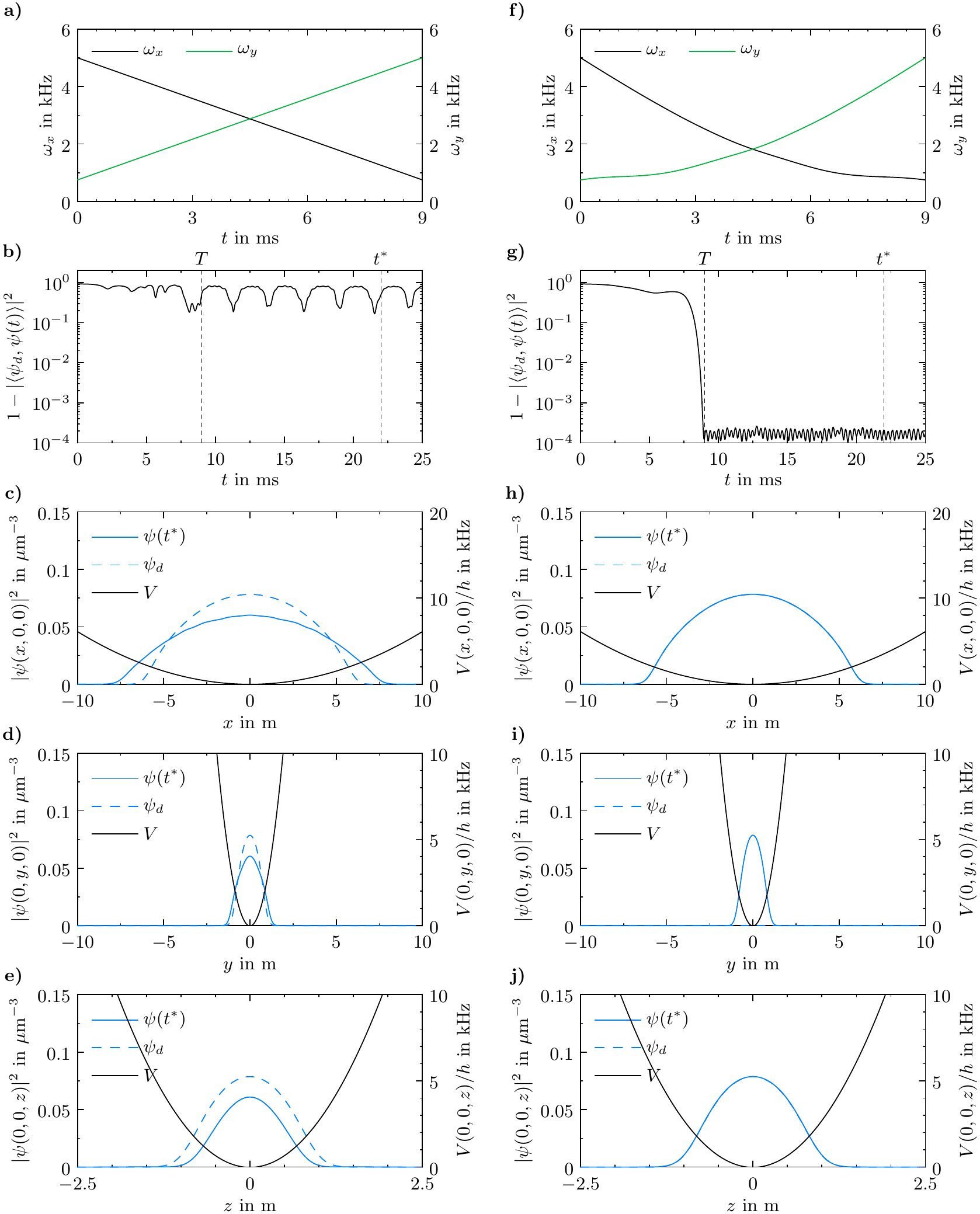}
	\caption{Two-parameter control of an elongated harmonic oscillator potential. 
	The computational domain is chosen as 
	$([-10,10] \times [-10,+10] \times [-2.5,2.5]) \, \mu \mathrm{m}^3$. 
	In the finest discretization level we use $128 \times 128 \times 32$ grid points and a 
	time step size of $\triangle t = 0.001$\,ms. 
	Left column: without optimal control. Right column: optimal control. For details see text. }
	\label{fig:harmonic_linear_versus_dual}
\end{figure*}

\subsubsection{Numerical simulations}

In the following simulations the number of atoms is $N=5000$, the final time is set to $T=9$\,ms and $\omega_z = 5$\,kHz.
The initial configuration of the trapping potential is given by
$\omega_x^i = 5$\,kHz and $\omega_y^i = 0.75$\,kHz, the final configuration by
$\omega_x^f = 0.75$\,kHz and $\omega_y^f = 5$\,kHz.

Before we discuss the result of the optimal control algorithm we first consider
a numerical simulation as a benchmark, in which the control parameters $\lambda_1$ and $\lambda_2$ are varied linearly. The corresponding time-evolution of the trap frequencies $\omega_x$ and $\omega_y$ is depicted in Fig.~\ref{fig:harmonic_linear_versus_dual}a. 

In order to investigate the overlap of $\psi$ with $\psi_d$ beyond the end of the control
we continue the time-evolution with $\blambda(t) = \blambda(T)$ for $t > T$. We proceed analogously in the other examples. As can be seen from Fig.~\ref{fig:harmonic_linear_versus_dual}b the infidelity decreases only slightly until $t=T$ and shows a strong oscillation for $t>T$. This behavior of the infidelity indicates that the final state differs significantly from the desired state $\psi_d$. This is also strikingly visualized by example snapshots of the density at time $t^* = 22$\,ms in Figs.~\ref{fig:harmonic_linear_versus_dual}c-e.

Next, we consider the result of the optimal control algorithm. 
Using $\blambda^0(t) = [0.25 \sin(\pi t / T) + t / T, -0.25 \sin(\pi t / T) + t / T]$ for $t \in [0,T]$ as a starting point,  
the algorithm converges to a solution
that reduces the cost functional by four orders of magnitude. 
The time-evolution of the frequencies $\omega_x$ and $\omega_y$ is shown in Fig.~\ref{fig:harmonic_linear_versus_dual}f, the time-evolution of the corresponding infidelity in Fig.~\ref{fig:harmonic_linear_versus_dual}g. It can clearly be seen that the infidelity strongly decreases until the end of the control at $t=T$. Moreover, the infidelity remains on a very low level for $t > T$, indicating that the desired final state has been reached with high precision. 
Consequently, the deviations of the density to the density of the desired state at time $t=t^*$ 
are very close to zero as can be seen from Figs.~\ref{fig:harmonic_linear_versus_dual}h-j. 
We note at this point that the evolution of the 3D wave functions can naturally only be described here in limited detail. 
A supplementary video that visualizes these dynamics in greater detail is available online~\cite{videos}.


\begin{figure*}[htb]
\centering
\includegraphics[width=0.925\textwidth]{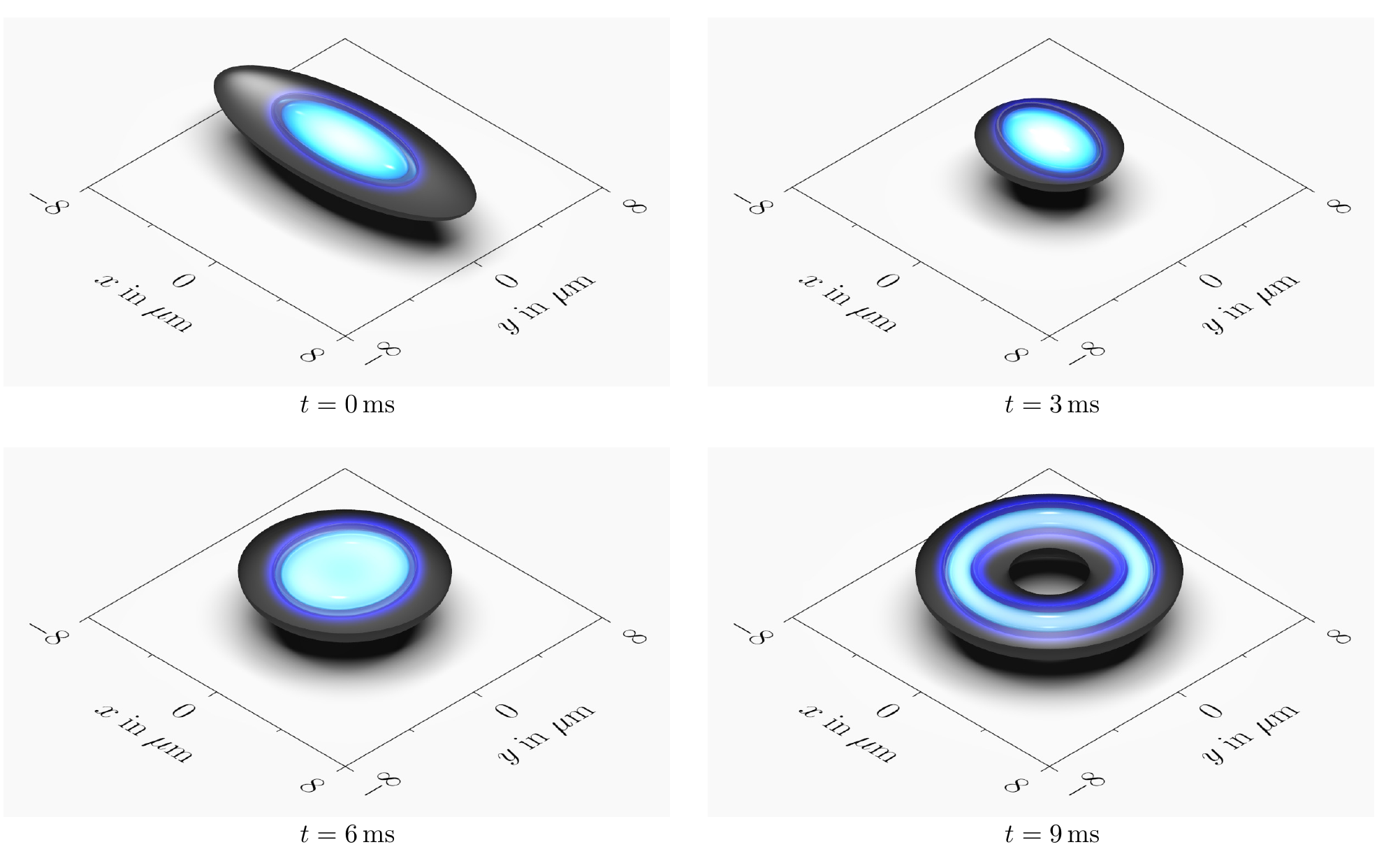}\\
\caption{Loading of a toroidal trap using two-parameter optimal control. Animations of the full dynamics are available online \cite{videos}. 
}
\label{fig:toroidal_snapshots_3d}
\end{figure*}

\begin{figure}[htb]
	\centering
	\includegraphics[width=0.45\textwidth]{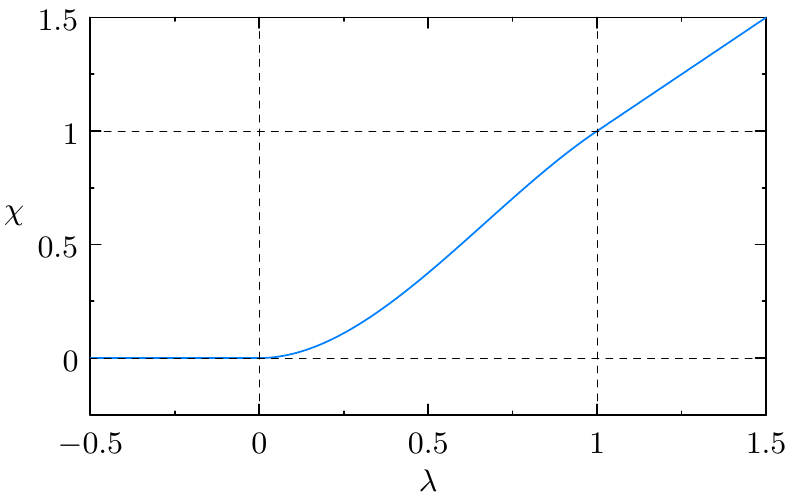}
	\caption{Saturation function used in the toroidal trap and splitting examples.}
	\label{fig:toroidal_chi_of_lambda}
\end{figure}

\subsection{Loading of a toroidal trap}
\label{sec:toroidal}
In the second application we consider the loading of a toroidal trap as shown in 
Fig.~\ref{fig:toroidal_snapshots_3d}. Such toroidal traps have recently been employed to realize atomic analogues of electrical circuits to study superflow and hysteresis~\cite{Ryu2007,Ryu2013,Beattie2013,Jendrzejewski2014,Eckel2014}. 

\subsubsection{Trapping potential}

The trapping potential is given by a slightly elongated harmonic potential and a Gaussian function centered at the origin of our coordinate system \cite{RyuAndCla2007}
\begin{equation*}
\label{eq:V_toroidal}
\begin{aligned}
V_\blambda (x,y,z) 
&= 
\frac{m}{2} 
\Big( 
\left[\omega_x(\lambda_1)\right]^2 x^2 + \omega_y^2 y^2 + \omega_z^2 z^2 \Big) \\
&\qquad + V_0(\lambda_2) \exp(-2 (x^2 + y^2) / w_0^2).
\end{aligned}
\end{equation*}
In an experiment this Gaussian function could for example correspond to a red-detuned laser beam realizing a repulsive dipole potential. 

As illustrated in Fig.~\ref{fig:toroidal_snapshots_3d} 
we consider the transformation of the potential
from an initial harmonic configuration with $\omega_x = \omega_x^i$ and $V_0 = 0$ at time $t=0$
to a toroidal configuration with $\omega_x = \omega_x^f$ and $V_0 = V_0^*$ at the final time $t=T$.
Hence, a suitable parameterization of $\omega_x$ and $V_0$ is given by
\begin{subequations}
\begin{align}
\label{eq:toroidal_omega_x}
\omega_{x}(\lambda_1) &= \omega_x^i + \lambda_1 (\omega_x^f - \omega_x^i),\\
\label{eq:toroidal_V_0}
V_0(\lambda_2) &= V_0^* \, \chi(\lambda_2),
\end{align}
\end{subequations}
where
\begin{align*}
\lambda_1(0) &= 0, \quad \lambda_1(T) = 1,\\
\lambda_2(0) &= 0, \quad \lambda_2(T) = 1.
\end{align*}
In Eq.~\eqref{eq:toroidal_V_0}, $\chi$ plays the role of a saturation function. The use of the saturation function ensures that $V_0$ remains positive - and thus experimentally realizable - for any possible choice of $\lambda_2$. This does not restrict the original control problem, as every experimentally realizable trajectory $V_0(t) \geq 0$ can be parametrized through a suitable $\lambda_2(t)$ in $V_0^* \chi(\lambda_2(t))$. In fact, we choose all parameters for the external potential to be close to previous experimental realizations. However, our approach also allows us to optimize more general situations where the parametrization of the trapping potential is more complicated~\cite{LesSchHoff2006}. 

Similar saturation functions are commonly used in control theory to realize limits on control parameters. 
In our particular case $\chi$ is implemented using a piecewise cubic hermite interpolating polynomial (PCHIP). 
Its functional form is shown in Fig.~\ref{fig:toroidal_chi_of_lambda}. 
The interpolating points are chosen such that $\chi$ always remains positive. 
Moreover, $\chi(0) = 0$ and $\chi(1) = 1$.

\subsubsection{Numerical simulations}

The following simulations are carried out using $V_0^* = h \times 30$\,kHz, $w_0 = 5\,\mu\mathrm{m}$
, $T=9$\,ms and $N=5000$. The frequencies
$\omega_y = 2.5$\,Hz and  $\omega_z = 5$\,kHz are kept constant during the simulation.
The initial configuration of the confinement potential is characterized by
$\omega_x^i = 1$\,kHz and $V_0(t=0) / h = 0$\,kHz, whereas the final
configuration is given by
$\omega_x^f = 2.5$\,kHz and $V_0(T)/h = V_0^*$.

\begin{figure*}[htb]
	\centering
	\includegraphics[width=0.965\textwidth]{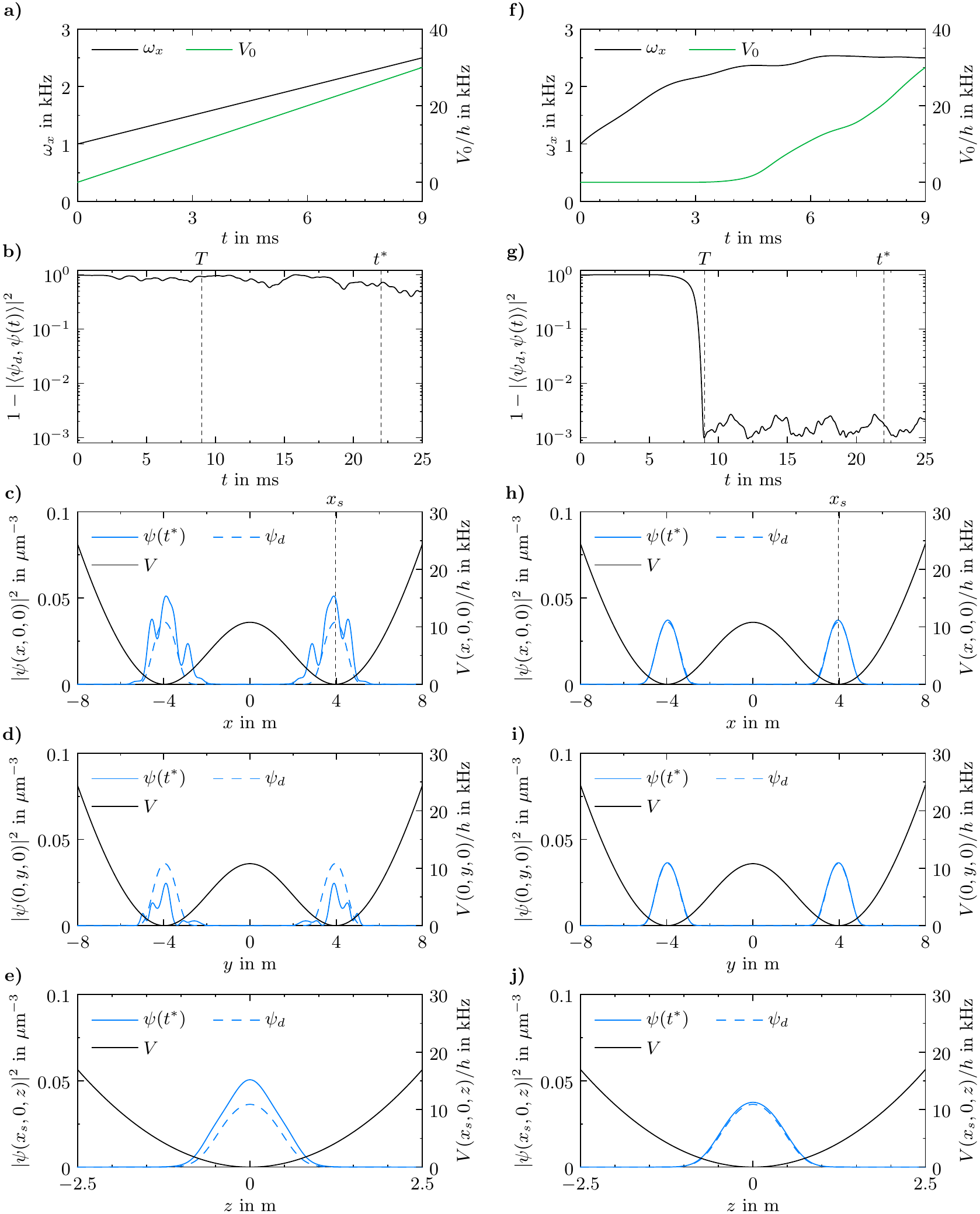}
	\caption{Loading of a toroidal trap using two-parameter control. 
	The computational domain is given by 
	$([-8,8] \times [-8,+8] \times [-2.5,2.5]) \, \mu \mathrm{m}^3$.  
	In the finest discretization level we use $128 \times 128 \times 40$ grid points
	and a time step size of $\triangle t = 0.001$\,ms. 
	Left column: linear variation of the control parameters. 
	Right column: optimal control of the control parameters. For details see text.}
	\label{fig:toroidal_linear_versus_dual}
\end{figure*}

As in the previous example we consider first the case where 
the parameters $\omega_x$ and $V_0$ are changed linearly (see Fig.~\ref{fig:toroidal_linear_versus_dual}a).
Fig.~\ref{fig:toroidal_linear_versus_dual}b reveals that the associated infidelity
does not drop at all until $t=T$. For $t > T$ we observe a slight decrease of the infidelity.
This can be attributed to the fact that, as time evolves, the density of the condensate becomes more evenly 
distributed in the toroidal trapping potential, bringing its wavefunction closer to $\psi_d$. 
However, as can be seen from Figs.~\ref{fig:toroidal_linear_versus_dual}c-e, the final wave function still differs strongly from the wave function of the desired state after $t^*=22\,$ms. 

Let us now discuss the result of the optimal control algorithm. An optimal time-evolution of the control parameters is given in Fig.~\ref{fig:toroidal_linear_versus_dual}f. Intuitively this control can be understood as the result of two separate time-scales. During the first halve of the control, the trap frequency $\omega_x$ is increased, while the limits imposed on $\lambda_2$ prohibit any change of $V_0$. During the second halve, on the other hand, $V_0$ is adjusted to its final value, while $\omega_x$ is only subject to small corrections. 

Until the end of the control this leads to a drop in the infidelity by approximately three orders of magnitude, as visualized in Fig.~\ref{fig:toroidal_linear_versus_dual}g. 
Furthermore, the infidelity remains bounded by $3 \times 10^{-3}$ for $t>T$, which is well below the measurement sensitivity in typical experiments. 
Consequently, only slight deviations from the desired wavefunction at time $t^* = 22$\,ms can be observed in Figs.~\ref{fig:toroidal_linear_versus_dual}h-j. 

\subsection{Splitting}
\label{sec:splitting}

\begin{figure*}[htb]
	\centering
	\includegraphics[width=0.925\textwidth]{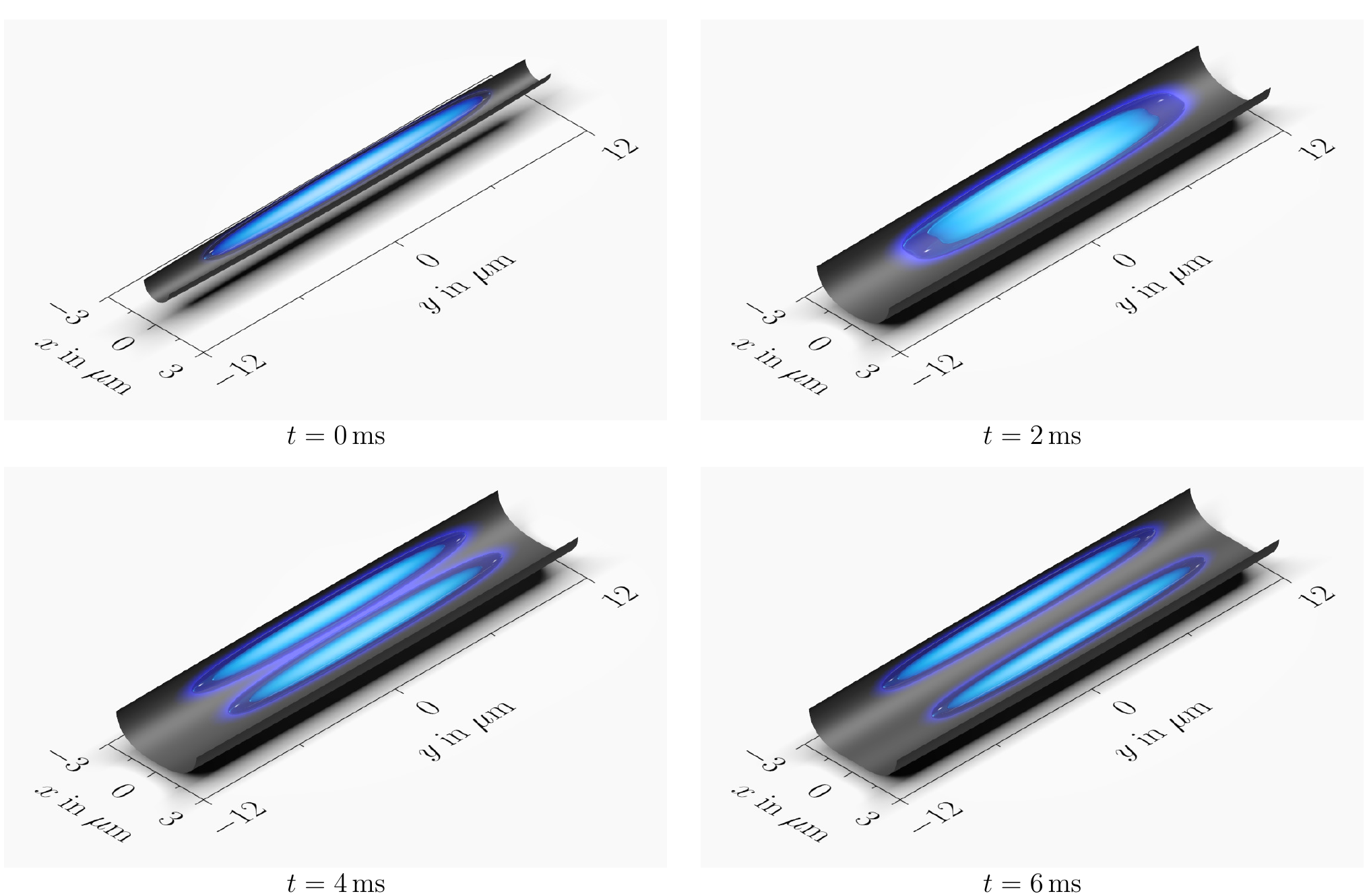}
	\caption{The splitting of a Bose-Einstein condensate, as realized by a radial deformation of an initially 
	harmonic potential into a double well~\cite{Schumm2005}. 
	The two gases in the final picture are completely decoupled, with no more overlap between the respective wave 
	functions. Animations of the full dynamics are available online \cite{videos}.}
	\label{fig:splitting_snapshots_3d}
\end{figure*}
In terms of technological applications, a particular noteworthy realization of BECs is achieved using atom chips~\cite{Folman2002,Reichel2010}. 
On these chips micro-fabricated wires allow the precise manipulation of BECs using static, radio and microwave fields. As a third application we thus consider the splitting of a single condensate into two identical halves using such an atom chip~\cite{Schumm2005}. 
A visualization is presented in Fig.~\ref{fig:splitting_snapshots_3d}. This splitting protocol has recently been used to study the non-equilibrium dynamics of 1D Bose gases, revealing subtle effects, such as prethermalization~\cite{Gring2012,Kuhnert2013,Smith2013,Langen2013b}, generalized statistical ensembles~\cite{Langen2015} and the light-cone-like emergence of thermal correlations~\cite{Langen2013,Geiger2014}. Moreover, it forms the basic building block for integrated matter-wave interferometers~\cite{Berrada2013,Grond2009}. 

\subsubsection{Trapping potential}

In the experiments the splitting is realized by dressing the static magnetic trapping potential with a strong near-field radio-frequency (RF) field. The unscaled static potential is given by $V_\mathrm{static} = g_F \mu_B m_F |\textbf{B}|$, with the magnetic field $\textbf{B}=(B_x,B_y,B_z)$ being well approximated by the famous Ioffe-Pritchard form
\begin{align*}
B_x &= B_1 x - \frac{B_2}{2}xy\\
B_z &= -B_1 z - \frac{B_2}{2}zy\\
B_y &= B_0+\frac{B_2}{2}\left[y^2-\frac{1}{2}\left(x^2+z^2\right)\right].
\end{align*}
The parameters are given by $B_0=\hbar \omega_\mathrm{0}/m_Fg_F\mu_B$, $B_1=\sqrt{m \omega_\perp^2 B_0/m_Fg_F\mu_B}$ and $B_2=m\omega_\parallel^2/m_Fg_F\mu_B$. 
In the following simulations we consider ${}^{87}$Rb atoms which are trapped in the $5\mathrm{S}_{1/2} \, F=2, m_F=2$ state where $g_F= 1/2$. 
The trap parameters are
\begin{align*}
\omega_\mathrm{0}                               &=  2 \pi \times 390\,\mathrm{kHz},\\
\omega_x = \omega_z \equiv \omega_\perp         &=  2 \pi \times   2\,\mathrm{kHz},\\
\omega_y \equiv \omega_\parallel                &=  2 \pi \times  85\,\mathrm{Hz}.
\end{align*}
The resulting dressed-state potential is given by~\cite{Lesanovski2006}
\begin{align*}
V_{\lambda} &=g_F \mu_B \tilde m_F \sqrt{\left(\frac{\hbar \omega_{RF}}{|g_F|\mu_B}-|\textbf{B}|\right)^2+\left(\frac{B_{RF\perp}(t)}{2}\right)^2}\\
&= g_F \mu_B \tilde m_F \sqrt{\Delta_\mathrm{RF}(\mathbf{r})^2+\Omega_\mathrm{rabi}^2(t)}
\end{align*}
with $\tilde m_F = 2$, $\omega_{RF}$ the frequency of the RF radiation and $B_{RF\perp}$ denoting the component of the linear polarized dressing field $\textbf{B}_{RF}$ that is aligned perpendicular to the static field. As in~\cite{Langen2013} we use a detuning of $\Delta_\mathrm{RF}(0) = - 2 \pi \times 30\,\mathrm{kHz}$ from the $m_F = 2 \rightarrow m_F = 1$ transition for the simulation. The Rabi-frequency is parameterized by the control parameter $\lambda$ 
\begin{equation*}
\label{eq:rabi_frequency}
\Omega_\mathrm{rabi}(\lambda) = \Omega_\mathrm{rabi}^* \, \chi(\lambda), \quad \Omega_\mathrm{rabi}^* = 2 \pi \times 155\,\mathrm{kHz}
\end{equation*}
wherein
\begin{align*}
\lambda_1(0) = 0, \quad \lambda_1(T) = 1.
\end{align*}
The control parameter $\lambda$ mimics the situation in experiments, where the double well potential is controlled by changing the RF field amplitude through an RF current in a wire. For $\lambda = 0$ we recover the static harmonic potential, whereas $\lambda=1$ corresponds to a fully separated double well with no wave function overlap between the two halves of the system. Since the Rabi-frequency is strictly positive in experiments we employ the same saturation function $\chi$ as in the previous example (cf. Fig.~\ref{fig:toroidal_chi_of_lambda}).

As the trapping potential is significantly changed during the splitting the atoms are radially displaced from their equilibrium position in the harmonic trap. Consequently, strong dipole and breathing oscillations are usually observed in experiments. 
This poses a strong limitation to the use of such systems as interferometers~\cite{Berrada2013}. 
The minimization of such excitations is therefore one of the main motivations for our optimization. 

\begin{figure*}[htb]
	\centering
	\includegraphics[width=0.965\textwidth]{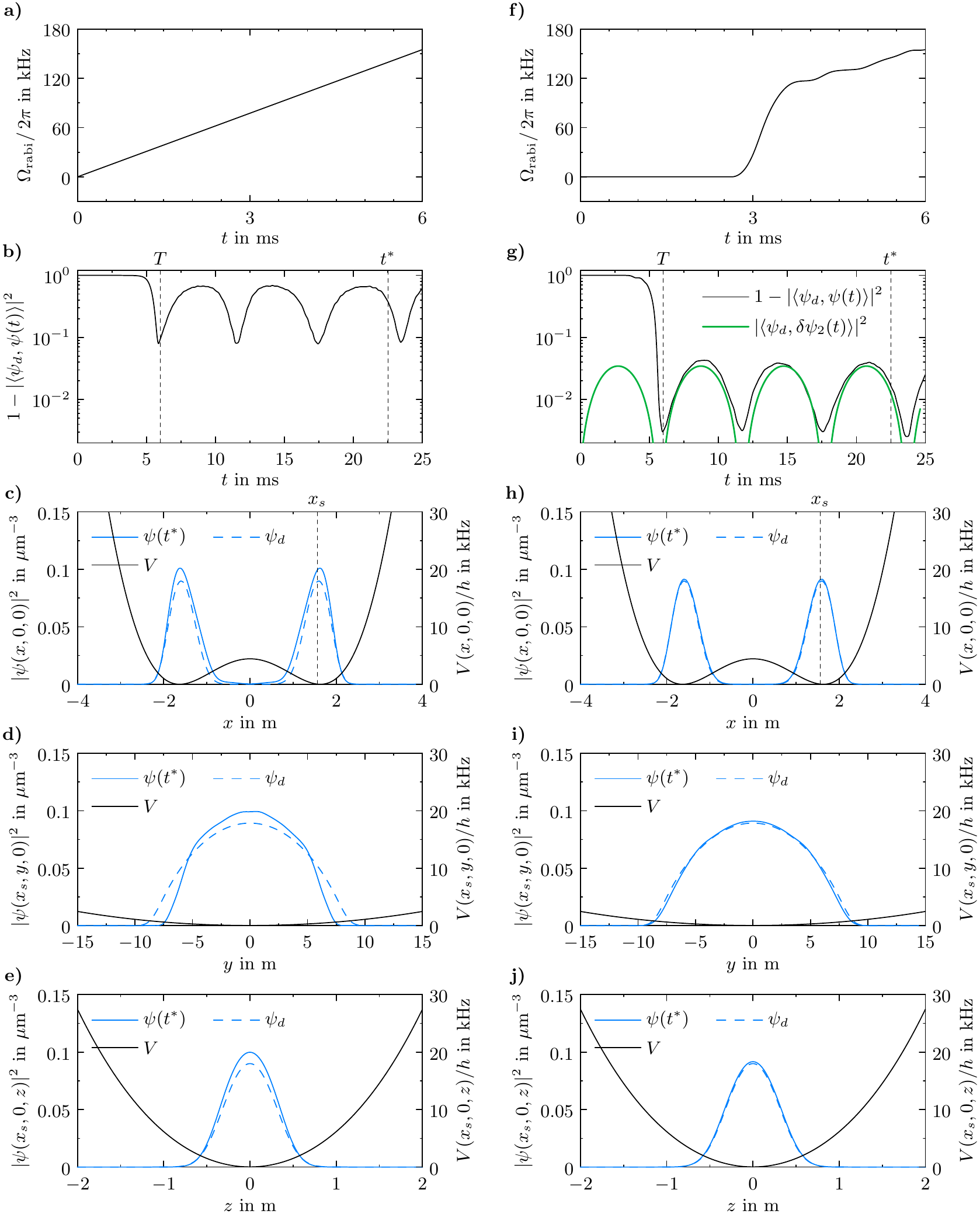}
	\caption{
	Splitting of a BEC using single-parameter optimal control. Left column: linear variation of the 
	control parameter.
	Right column: optimal control of the control parameter. 
	We note that the $|\langle \psi_d, \delta \psi_2(t) \rangle|^2$ in (g) has been scaled and 
	slightly shifted in time to account for the unknown phase and amplitude of the excitation. 
	The computational domain is given by  
	$([-4,4] \times [-15,+15] \times [-2,2]) \, \mu \mathrm{m}^3$
	which is discretized by $96\times 128 \times 48$ grid points in the finest discretization level.
	The corresponding time step is $\triangle t = 0.001$\,ms.  
	For details see text.
	}
	\label{fig:splitting_linear}
\end{figure*}

\subsubsection{Numerical simulations: single-parameter control}

We illustrate the splitting procedure for $N=2000$ atoms and $T=6$\,ms.

In a first step we again consider the case where the Rabi-frequency is increased linearly
(see Fig.~\ref{fig:splitting_linear}a). This procedure is identical to the one that is typically used in experiments \cite{Langen2015,Gring2012}. At the final time $t=T$ the infidelity has only decreased slightly as can be seen from Fig.~\ref{fig:splitting_linear}b. Moreover, the infidelity shows the expected strong oscillations for $t>T$.
A snapshot of the density at time $t^*=22.5$\,ms is illustrated in Figs.~\ref{fig:splitting_linear}c-e, revealing that there is large discrepancy between the computed state $\psi$ and the desired state $\psi_d$.

Next, we consider the result of the optimal control algorithm. 
We find that, irrespective of the specific choice of $\lambda^0$, the algorithm always converges to approximately the same minimizer of the cost functional. The corresponding time-evolution of the Rabi-frequency is shown in Fig.~\ref{fig:splitting_linear}f.
We observe that the Rabi-frequency remains zero for the first few milliseconds. In fact, only about three milliseconds of the optimization time $T$ are used for the transformation of the external potential. This behavior persists even if we increase the optimization time $T$, with the Rabi-frequency vanishing for an even longer initial period of time. The precise timescale depends on the parameters of the trap, as the optimization algorithm tries to find a compromise between longitudinal and radial directions. 

Interestingly, our 3D control qualitatively resembles the result of a previous 1D optimization that included beyond mean-field effects to model the distribution of atoms into the two final gases on the quantum level~\cite{Grond2009}. 
In both cases, the initial BEC is first rapidly split into two halves. 
Subsequently, these two halves are kept close enough to experience a tunnel coupling for a finite time-scale.
This qualitative observation is very interesting, as reducing relative number fluctuations can help to significantly enhance the sensitivity of such interferometers. 
A detailed study of how useful our control can be in this context will be a natural extension of this work. 

As a result of the optimal control algorithm the infidelity at the final time $T$ is reduced by more than two orders of magnitude
(see Fig.~\ref{fig:splitting_linear}g). 
However, for $t>T$ we again observe a strong oscillation.
Snapshots of the density distribution at $t^*=22.5$\,ms are given in
Figs.~\ref{fig:splitting_linear}h-j.

\subsubsection{Bogoliubov-de Gennes analysis}

\begin{figure*}[htb]
	\centering
		\includegraphics[width=0.925\textwidth]{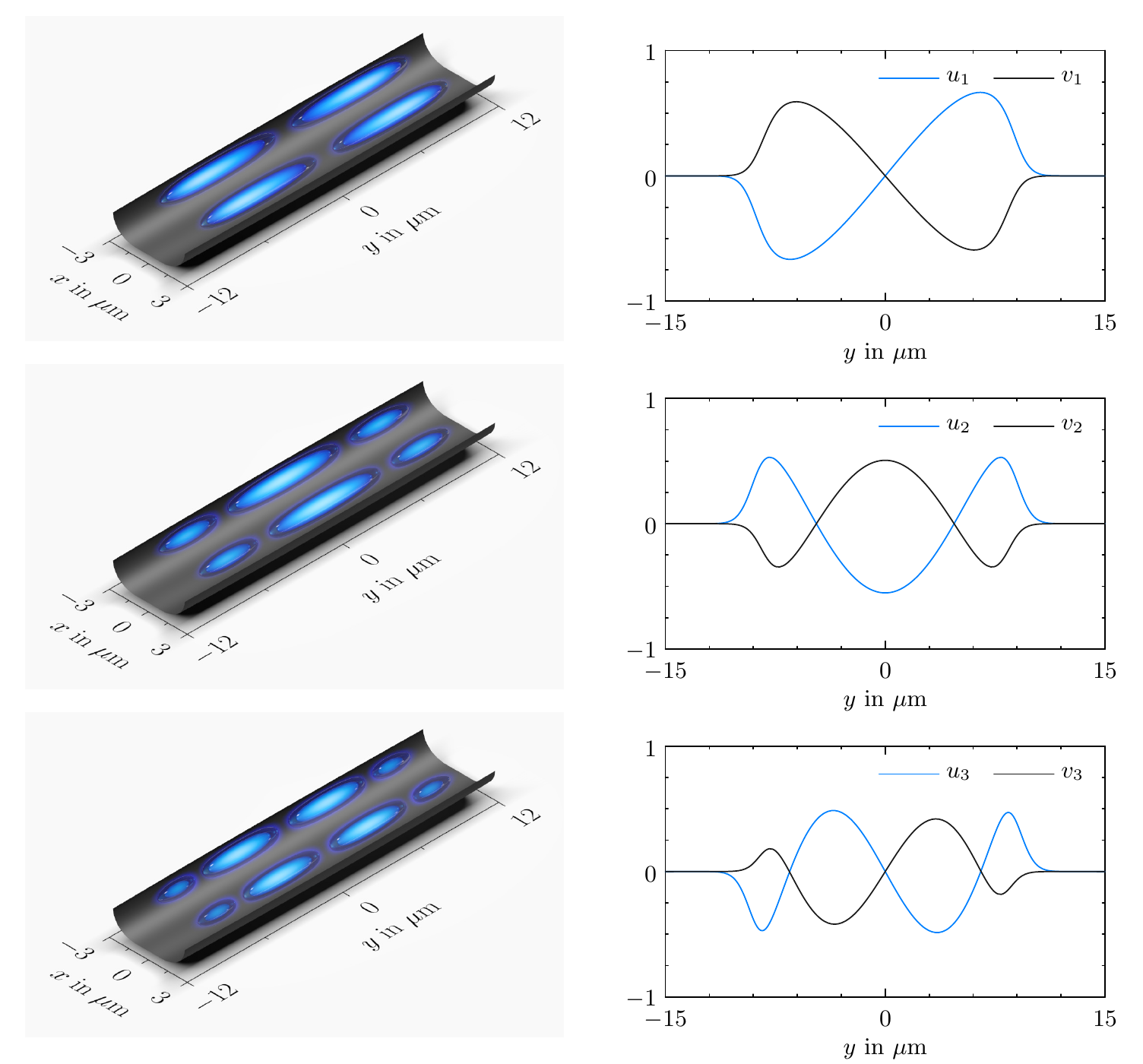}
  		\caption{
  		Solutions of the Bogoliubov-de Gennes equations using a 6th-order finite difference discretization for 
  		$N=2000$ atoms.
		Left: Density of the first three (scaled) excitations 
		$\delta \psi_1(\bm{r},t)$, $\delta \psi_2(\bm{r},t)$
		and $\delta \psi_3(\bm{r},t)$ 
		at $t = \mathcal{T}_{\mathrm{eff},1}/2$,
		$t = \mathcal{T}_{\mathrm{eff},2}/2$ and $t = \mathcal{T}_{\mathrm{eff},3}/2$.
		Right: Normalized (with respect to the inner product~\eqref{eq:norm_bdg}) amplitude 
		functions $u$ and $v$ evaluated along the longitudinal direction at $x=x_s$ and $z=0$.
		All functions are purely real-valued.
		}
\label{fig:vsi_solutions_BdG}
\end{figure*}

Interestingly, the $6$\,ms period of the very regular infidelity oscillation shown in Fig.~\ref{fig:splitting_linear}g for the optimized splitting is approximately the same as the period of the infidelity oscillation depicted in Fig.~\ref{fig:splitting_linear}b for the simple linear splitting. This suggests that the character of the oscillation is determined by the intrinsic properties of the BEC rather than by the splitting protocol.

Indeed, we demonstrate in the following that the oscillations are caused by collective excitations of the BEC, which are created during, but irrespective of the details of the splitting process. To this end, we show that they are the result of a small deviation $\delta \psi$ from the desired state $\psi_d$, which can be described within the Bogoliubov-de Gennes (BdG) framework. 

Let therefore $\Phi(\bm{r},t) = \phi(\bm{r}) e^{-i \mu t / \hbar}$ denote an 
eigenstate solution of the GPE. 
Here, $\mu$ is the corresponding chemical potential and $\phi$ is a solution of the stationary GPE, $H_0 \phi + g |\phi|^2 \phi = \mu \phi$, with $H_0 = -\hbar^2 / 2 m \, \Delta + V$. 
We consider a generic state $\psi$ which deviates from the eigenstate solution by a small 
fluctuation $\delta \psi$, i.e,
\begin{equation}
\label{eq:perturbation_psi}
\psi(\bm{r},t) \approx \Phi(\bm{r},t) + \delta \psi(\bm{r},t).
\end{equation}
In a linear approximation (with respect to $\delta \psi$) this small deviation is given by
\begin{equation}
\label{eq:delta_psi}
\delta \psi(\bm{r},t) = \big( u(\bm{r}) e^{-i \omega t} + v^*(\bm{r}) e^{i \omega^* t} \big) e^{-i \mu t / \hbar}
\end{equation}
where $u$, $v$ and $\omega$ are defined via the solutions of the BdG equations \cite{Le2001,TaNaYa2014}
\begin{equation}
\label{eq:BdG}
\begin{bmatrix}
H_0 - \mu + 2 g |\phi|^2 & g \phi^2 \\
-(g \phi^2)^* & -H_0 + \mu - 2 g |\phi|^2
\end{bmatrix}
\begin{bmatrix}
u \\
v
\end{bmatrix}
= 
\hbar \omega
\begin{bmatrix}
u \\
v
\end{bmatrix}.
\end{equation}

We want to investigate small fluctuations $\delta \psi$ corresponding to some of the lowest energy eigenvalues $\hbar \omega$ in equation \eqref{eq:BdG}. To this end, we proceed in a conceptually similar way to \cite{HuTuMeBra2003} where numerical methods are used to investigate the stability and decay rates of non-isotropic attractive Bose-Einstein condensates. Like in Ref. \cite{HuTuMeBra2003} we consider the full three-dimensional problem.
However, for the discretization of the operators in~\eqref{eq:BdG} we employ a high-order finite difference discretization rather than working in a Fourier basis. By gradually increasing the spatial resolution of the finite difference discretization we are able to verify the convergence of the algorithm. A detailed description of our implementation is again given in the Appendix.

As an example, we find that the first three eigenvalues converge towards
$\omega_1 = \pm 314.54$\,Hz, 
$\omega_2 = \pm 523.49$\,Hz and 
$\omega_3 = \pm 734.26$\,Hz.
Subsequently, the corresponding eigenfunctions $(u_i, v_i)$ are normalized according to the norm \cite{TaNaYa2014}
\begin{equation}
\label{eq:norm_bdg}
\int_{\mathbbm{R}^3} \big( u_i^2(\bm{r}) - v_i^2({\bm{r}})\big)\, d\bm{r} = 1.
\end{equation}

Knowing the frequencies $\omega_i$ and amplitude functions $u_i$ and $v_i$, it is possible to
investigate the time-evolution of the excitations given by Eq.~\eqref{eq:delta_psi}.
It turns out that $|\delta\psi_i(t)|^2$ can be well described by a simple periodic oscillation in amplitude, while the shape remains mostly
unchanged (see left column in Fig.~\ref{fig:vsi_solutions_BdG}). 
As $u_i$ and $v_i$ are purely real-valued functions, which approximately fulfill $v_i = -u_i$ (see right column in Fig.~\ref{fig:vsi_solutions_BdG}) we find
$\delta \psi_i(\br,\mathcal{T}_i/2) \approx \delta \psi_i(\br, \mathcal{T}_i)$
and hence the effective oscillation periods are halved
with respect to the eigenvalues found above, i.e. 
$\mathcal{T}_{\mathrm{eff},i} = \mathcal{T}_i / 2=\pi/\omega_i$. In detail we find $\mathcal{T}_{\mathrm{eff},1} = 9.99$\,ms, $\mathcal{T}_{\mathrm{eff},2} = 6.00$\,ms and $\mathcal{T}_{\mathrm{eff},3} = 4.28$\,ms. 

Note that the effective period of the second excitation is very close to the period of the oscillation of the infidelity observed above. Indeed, plotting the time-evolution of $|\langle \psi_d, \delta \psi_2(t) \rangle|^2$ along with the time-evolution of the infidelity in Fig.~\ref{fig:splitting_linear}g demonstrates clearly that the oscillation of the infidelity is dominated by the second excitation. As further evidence, we extract the deviation of $\psi$ from $\psi_d$ from our simulation. A comparison shows again very good agreement with the time-evolution of $\delta\psi_2(t)$ (see Appendix).

The fact that only the second but not the first excitation contributes to the observations can be understood from symmetry arguments. The first excitation corresponds to an antisymmetric wave function with respect to the 
longitudinal direction, whereas the second excitation is symmetric. During the splitting process, the halving of the atom number in each of the two gases, as well as an overall change in the longitudinal trapping potential leads to a symmetric change in the extension of the BEC in this direction. If the control is unable to compensate for this change in extension, the second Bogoliubov-de Gennes mode is automatically excited. 


This effect is especially pronounced for the linear splitting. 
In contrast to that, the optimal control algorithm can still reduce the infidelity at $t=T$, but 
even a small deviation of the wave function from the stationary state leads
to a strong oscillation in the infidelity for $t \geq T$.

Once the wave function differs from the stationary state in the longitudinal direction it is impossible to stop the observed oscillation by a simple variation of the Rabi-frequency. The BEC will thus oscillate for $t>T$ after the end of the control.

A central role in this scenario is played by the longitudinal frequency $\omega_y$. The smaller $\omega_y$ the longer the extension of the condensate in the longitudinal direction. In analogy to a classical harmonic oscillator this increases the susceptibility  to small deviations from the equilibrium position. 
We have confirmed this intuition with additional simulations, finding an even more pronounced excitation of the second mode for smaller $\omega_y$.

This is particularly noteworthy with respect to experiments studying BECs in the one-dimensional limit, where $\omega_{x,z}\gg\omega_y$~\cite{Langen2015}. 
Intuitively, such experiments should be very well described through a 1D approximation, where only a reduced GPE for the $x$-direction has to be considered (see Appendix). Our results here show that such an approach will, in general, also lead to a strong breathing oscillation. Even if the 1D control is able to reach the 1D desired state with high precision, it does not necessarily describe the experimental reality and will thus fail in 3D.

\subsubsection{Numerical simulations: two-parameter control}

\begin{figure*}[htb]
	\centering
	\includegraphics[width=0.975\textwidth]{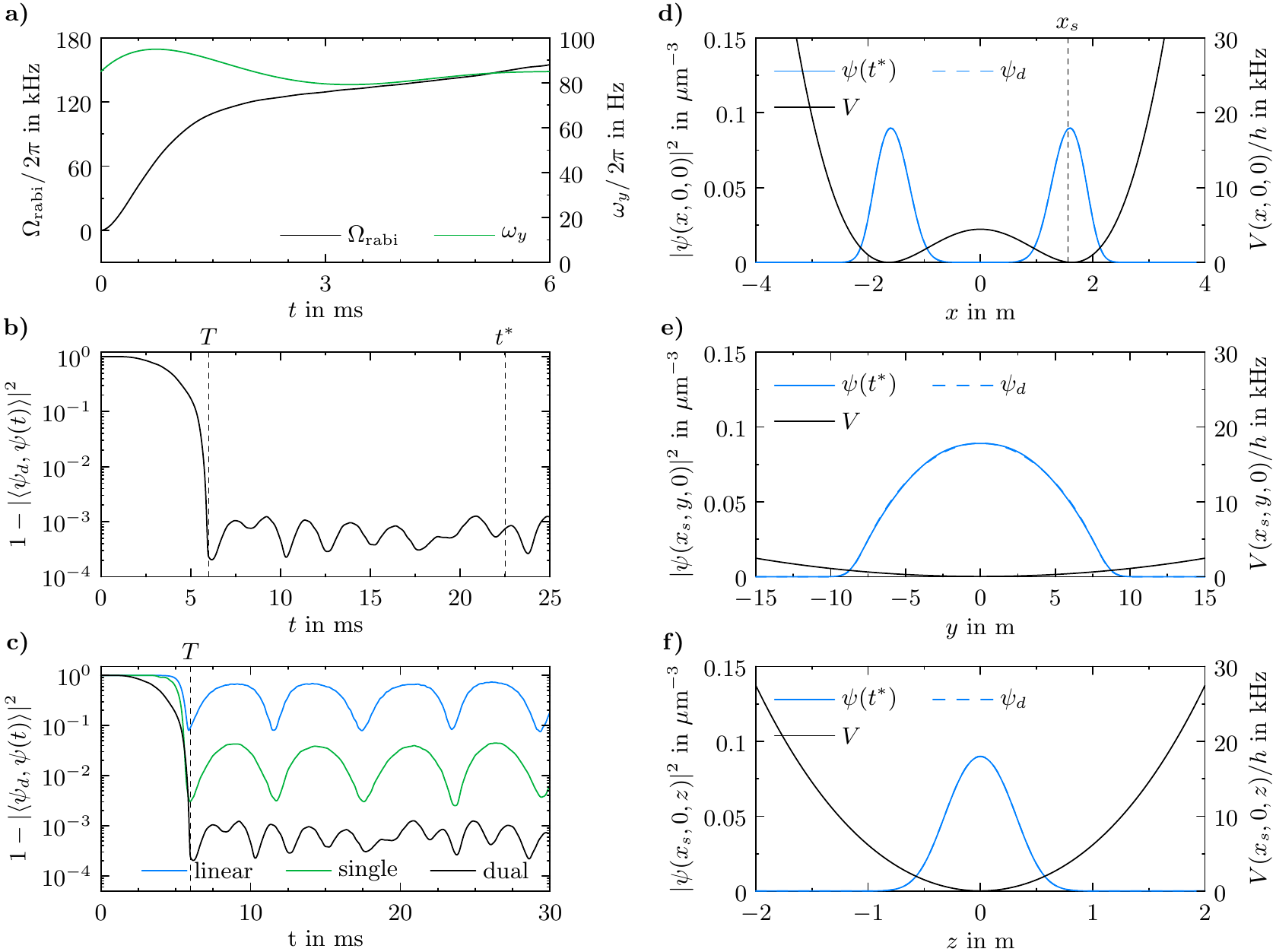}
	\caption{(a) Time-evolution of $\Omega_\mathrm{rabi}$ and $\omega_y$ corresponding to two-parameter 
	optimal control of the splitting process.
	(b) Associate infidelity.
	(c) Comparison of the infidelities for the linear and the optimal single-
	and two-parameter control.
	(d),(e) and (f) Snapshots of the density at time $t^* = 22.5$\,ms.}
	\label{fig:splitting_double}
\end{figure*}

In the last part of this article we will show how the oscillations reported above can be eliminated
using a more sophisticated control scheme that is made possible by the 3D character of our control and that involves a manipulation of the trapping potential along the longitudinal direction. In experiments on atom chips, this manipulation can, for example, be realized using additional wire structures, which provide longitudinal confinement independent of the main radial trapping structures~\cite{LangenThesis}. 

In analogy to the previous examples, we consider the following parameterization of $\Omega_\mathrm{rabi}$ and $\omega_y$:
\begin{subequations}
\begin{align}
\Omega_\mathrm{rabi}(\lambda_1) &= \Omega_\mathrm{rabi}^* \, \chi(\lambda_1), &  \Omega_\mathrm{rabi}^*  &= 2 \pi \times 155\,\mathrm{kHz}, \\
\omega_y(\lambda_2)             &= \omega_y^* \, \lambda_2,                   &  \omega_y^*              &= 2 \pi \times  85\,\mathrm{Hz},
\end{align}
\end{subequations}
with
\begin{align*}
\lambda_1(0) &= 0, \quad \lambda_1(T) = 1,\\
\lambda_2(0) &= 1, \quad \lambda_2(T) = 1.
\end{align*}
The only difference to the previous example is thus that the value of the longitudinal trap frequency $\omega_y$ is now part of the control. We still fix $\omega_y(t=0) = 2 \pi \times 85$\,Hz and $\omega_y(t=T) = 2 \pi \times 85$\,Hz such that the initial and desired final states remain unchanged. 

Using $\blambda^{0}(t) = [t / T, 1]$ for $t \in [0,T]$ as an initial guess the optimization algorithm
converges to a solution which reduces the cost functional by more than three orders of magnitude.
The time-evolution of the corresponding physical parameters is given in Fig.~\ref{fig:splitting_double}a.
As can be seen from Fig.~\ref{fig:splitting_double}b the infidelity remains very low for $t \geq T$. 
Snapshots of the density distribution at time $t^\star=22.5$\,ms confirm that the deviation from the desired state is extremely small, see Figs.~\ref{fig:splitting_double}d-f.

In the given example we have chosen $T = \mathcal{T}_{\mathrm{eff},2}$. In contrast to that, for a time $T<\mathcal{T}_{\mathrm{eff},2}$ we find significantly worse results. The minimum time scale $T$ is thus set by the oscillation period of the excitation that the control aims to stop. This oscillation period is in turn set by the geometry of the trap. Each different experimental situation will thus require carefully chosen parameters for the control.


\section{Conclusion and outlook}

In this work we have presented the first optimal control of the GPE in 3D. As we have shown, this situation is inherently more difficult than the optimal control of the 1D GPE because of the non-linear coupling of different coordinate directions. We have performed a detailed analysis of the resulting small excitations, which we were able to minimize by extending previous control schemes from a single to a multi-parameter control. 

In contrast to 1D approximations our 3D approach
allows the study of realistic trapping potentials, which will have direct impact on the quality of experiments and therefore provide an important step in the ongoing effort to use the properties of the quantum world for real life applications. Importantly, our scheme is not limited to the examples discussed in this work but rather very flexible, with many more applications conceivable. 

A straight-forward extension of our numerical solver could include the treatment of excited states. This would allow the three-dimensional study of a recent experiment, where the BEC was transferred to the first excited state of the trapping potential via a 1D optimal control sequence~\cite{Bucker2011}. Based on our observations we expect an even stronger excitation of BdG-Modes in such an experiment. In that context, another interesting application would be to replace the cigar-shaped confinement potentials used in the splitting and vibrational state inversion experiments by torus-shaped trapping potentials. Due to the different topology the issues related to the excitation of small perturbations are expected to be strongly reduced.

Another obvious extension of this work could be to consider different cost functionals. 
More precisely, it would be interesting to investigate whether it is possible to reduce the optimization times $T$ by using other cost functionals which are not based on the infidelity but rather on a  conserved quantity like the total energy.

Finally, interesting further directions include the study of beyond mean-field effects using the multi-configurational time-dependent Hartree framework for bosons~\cite{Meyer1990} or the optimization of finite temperature states. 


\section{Acknowledgments}
We thank Jörg Schmiedmayer, Wolfgang Rohringer and Alexander Pikovski for helpful discussions. R.M.L. is supported by the Hertha-Firnberg Program of the Austrian Science Fund (FWF), Grant T402-N13. D.M. acknowledges Deutsche Forschungsgemeinschaft Collaborative Research Center TRR 109, “Discretization in Geometry and Dynamics”.  T.L. acknowledges support by the FWF through the Doctoral Programme CoQuS (W1210) and by the Alexander von Humboldt Foundation through a Feodor Lynen Research Fellowship. 

\bibliography{biblio}

\begin{thebibliography}{69}
\expandafter\ifx\csname natexlab\endcsname\relax\def\natexlab#1{#1}\fi
\expandafter\ifx\csname bibnamefont\endcsname\relax
  \def\bibnamefont#1{#1}\fi
\expandafter\ifx\csname bibfnamefont\endcsname\relax
  \def\bibfnamefont#1{#1}\fi
\expandafter\ifx\csname citenamefont\endcsname\relax
  \def\citenamefont#1{#1}\fi
\expandafter\ifx\csname url\endcsname\relax
  \def\url#1{\texttt{#1}}\fi
\expandafter\ifx\csname urlprefix\endcsname\relax\def\urlprefix{URL }\fi
\providecommand{\bibinfo}[2]{#2}
\providecommand{\eprint}[2][]{\url{#2}}

\bibitem[{\citenamefont{Bloch et~al.}(2008)\citenamefont{Bloch, Dalibard, and
  Zwerger}}]{Bloch08}
\bibinfo{author}{\bibfnamefont{I.}~\bibnamefont{Bloch}},
  \bibinfo{author}{\bibfnamefont{J.}~\bibnamefont{Dalibard}}, \bibnamefont{and}
  \bibinfo{author}{\bibfnamefont{W.}~\bibnamefont{Zwerger}},
  \bibinfo{journal}{Reviews of Modern Physics} \textbf{\bibinfo{volume}{80}},
  \bibinfo{pages}{885} (\bibinfo{year}{2008}).

\bibitem[{\citenamefont{Gross et~al.}(2010)\citenamefont{Gross, Zibold,
  Nicklas, Esteve, and Oberthaler}}]{Gross2010}
\bibinfo{author}{\bibfnamefont{C.}~\bibnamefont{Gross}},
  \bibinfo{author}{\bibfnamefont{T.}~\bibnamefont{Zibold}},
  \bibinfo{author}{\bibfnamefont{E.}~\bibnamefont{Nicklas}},
  \bibinfo{author}{\bibfnamefont{J.}~\bibnamefont{Esteve}}, \bibnamefont{and}
  \bibinfo{author}{\bibfnamefont{M.~K.} \bibnamefont{Oberthaler}},
  \bibinfo{journal}{Nature} \textbf{\bibinfo{volume}{464}},
  \bibinfo{pages}{1165} (\bibinfo{year}{2010}).

\bibitem[{\citenamefont{L{\"u}cke et~al.}(2011)\citenamefont{L{\"u}cke,
  Scherer, Kruse, Pezz{\'e}, Deuretzbacher, Hyllus, Peise, Ertmer, Arlt, Santos
  et~al.}}]{Lucke2011}
\bibinfo{author}{\bibfnamefont{B.}~\bibnamefont{L{\"u}cke}},
  \bibinfo{author}{\bibfnamefont{M.}~\bibnamefont{Scherer}},
  \bibinfo{author}{\bibfnamefont{J.}~\bibnamefont{Kruse}},
  \bibinfo{author}{\bibfnamefont{L.}~\bibnamefont{Pezz{\'e}}},
  \bibinfo{author}{\bibfnamefont{F.}~\bibnamefont{Deuretzbacher}},
  \bibinfo{author}{\bibfnamefont{P.}~\bibnamefont{Hyllus}},
  \bibinfo{author}{\bibfnamefont{J.}~\bibnamefont{Peise}},
  \bibinfo{author}{\bibfnamefont{W.}~\bibnamefont{Ertmer}},
  \bibinfo{author}{\bibfnamefont{J.}~\bibnamefont{Arlt}},
  \bibinfo{author}{\bibfnamefont{L.}~\bibnamefont{Santos}},
  \bibnamefont{et~al.}, \bibinfo{journal}{Science}
  \textbf{\bibinfo{volume}{334}}, \bibinfo{pages}{773} (\bibinfo{year}{2011}).

\bibitem[{\citenamefont{Riedel et~al.}(2010)\citenamefont{Riedel, B{\"o}hi, Li,
  H{\"a}nsch, Sinatra, and Treutlein}}]{Riedel2010}
\bibinfo{author}{\bibfnamefont{M.~F.} \bibnamefont{Riedel}},
  \bibinfo{author}{\bibfnamefont{P.}~\bibnamefont{B{\"o}hi}},
  \bibinfo{author}{\bibfnamefont{Y.}~\bibnamefont{Li}},
  \bibinfo{author}{\bibfnamefont{T.~W.} \bibnamefont{H{\"a}nsch}},
  \bibinfo{author}{\bibfnamefont{A.}~\bibnamefont{Sinatra}}, \bibnamefont{and}
  \bibinfo{author}{\bibfnamefont{P.}~\bibnamefont{Treutlein}},
  \bibinfo{journal}{Nature} \textbf{\bibinfo{volume}{464}},
  \bibinfo{pages}{1170} (\bibinfo{year}{2010}).

\bibitem[{\citenamefont{Ockeloen et~al.}(2013)\citenamefont{Ockeloen, Schmied,
  Riedel, and Treutlein}}]{Ockeloen2013}
\bibinfo{author}{\bibfnamefont{C.~F.} \bibnamefont{Ockeloen}},
  \bibinfo{author}{\bibfnamefont{R.}~\bibnamefont{Schmied}},
  \bibinfo{author}{\bibfnamefont{M.~F.} \bibnamefont{Riedel}},
  \bibnamefont{and}
  \bibinfo{author}{\bibfnamefont{P.}~\bibnamefont{Treutlein}},
  \bibinfo{journal}{Physical review letters} \textbf{\bibinfo{volume}{111}},
  \bibinfo{pages}{143001} (\bibinfo{year}{2013}).

\bibitem[{\citenamefont{Wildermuth et~al.}(2005)\citenamefont{Wildermuth,
  Hofferberth, Lesanovsky, Haller, Andersson, Groth, Bar-Joseph, Kr{\"u}ger,
  and Schmiedmayer}}]{Wildermuth05}
\bibinfo{author}{\bibfnamefont{S.}~\bibnamefont{Wildermuth}},
  \bibinfo{author}{\bibfnamefont{S.}~\bibnamefont{Hofferberth}},
  \bibinfo{author}{\bibfnamefont{I.}~\bibnamefont{Lesanovsky}},
  \bibinfo{author}{\bibfnamefont{E.}~\bibnamefont{Haller}},
  \bibinfo{author}{\bibfnamefont{L.~M.} \bibnamefont{Andersson}},
  \bibinfo{author}{\bibfnamefont{S.}~\bibnamefont{Groth}},
  \bibinfo{author}{\bibfnamefont{I.}~\bibnamefont{Bar-Joseph}},
  \bibinfo{author}{\bibfnamefont{P.}~\bibnamefont{Kr{\"u}ger}},
  \bibnamefont{and}
  \bibinfo{author}{\bibfnamefont{J.}~\bibnamefont{Schmiedmayer}},
  \bibinfo{journal}{Nature} \textbf{\bibinfo{volume}{435}},
  \bibinfo{pages}{440} (\bibinfo{year}{2005}).

\bibitem[{\citenamefont{Aigner et~al.}(2008)\citenamefont{Aigner, Della~Pietra,
  Japha, Entin-Wohlman, David, Salem, Folman, and Schmiedmayer}}]{Aigner08}
\bibinfo{author}{\bibfnamefont{S.}~\bibnamefont{Aigner}},
  \bibinfo{author}{\bibfnamefont{L.}~\bibnamefont{Della~Pietra}},
  \bibinfo{author}{\bibfnamefont{Y.}~\bibnamefont{Japha}},
  \bibinfo{author}{\bibfnamefont{O.}~\bibnamefont{Entin-Wohlman}},
  \bibinfo{author}{\bibfnamefont{T.}~\bibnamefont{David}},
  \bibinfo{author}{\bibfnamefont{R.}~\bibnamefont{Salem}},
  \bibinfo{author}{\bibfnamefont{R.}~\bibnamefont{Folman}}, \bibnamefont{and}
  \bibinfo{author}{\bibfnamefont{J.}~\bibnamefont{Schmiedmayer}},
  \bibinfo{journal}{Science} \textbf{\bibinfo{volume}{319}},
  \bibinfo{pages}{1226} (\bibinfo{year}{2008}).

\bibitem[{\citenamefont{Geiger et~al.}(2011)\citenamefont{Geiger, M{\'e}noret,
  Stern, Zahzam, Cheinet, Battelier, Villing, Moron, Lours, Bidel
  et~al.}}]{Geiger2011}
\bibinfo{author}{\bibfnamefont{R.}~\bibnamefont{Geiger}},
  \bibinfo{author}{\bibfnamefont{V.}~\bibnamefont{M{\'e}noret}},
  \bibinfo{author}{\bibfnamefont{G.}~\bibnamefont{Stern}},
  \bibinfo{author}{\bibfnamefont{N.}~\bibnamefont{Zahzam}},
  \bibinfo{author}{\bibfnamefont{P.}~\bibnamefont{Cheinet}},
  \bibinfo{author}{\bibfnamefont{B.}~\bibnamefont{Battelier}},
  \bibinfo{author}{\bibfnamefont{A.}~\bibnamefont{Villing}},
  \bibinfo{author}{\bibfnamefont{F.}~\bibnamefont{Moron}},
  \bibinfo{author}{\bibfnamefont{M.}~\bibnamefont{Lours}},
  \bibinfo{author}{\bibfnamefont{Y.}~\bibnamefont{Bidel}},
  \bibnamefont{et~al.}, \bibinfo{journal}{Nature communications}
  \textbf{\bibinfo{volume}{2}}, \bibinfo{pages}{474} (\bibinfo{year}{2011}).

\bibitem[{\citenamefont{Bloom et~al.}(2014)\citenamefont{Bloom, Nicholson,
  Williams, Campbell, Bishof, Zhang, Zhang, Bromley, and Ye}}]{Bloom14}
\bibinfo{author}{\bibfnamefont{B.}~\bibnamefont{Bloom}},
  \bibinfo{author}{\bibfnamefont{T.}~\bibnamefont{Nicholson}},
  \bibinfo{author}{\bibfnamefont{J.}~\bibnamefont{Williams}},
  \bibinfo{author}{\bibfnamefont{S.}~\bibnamefont{Campbell}},
  \bibinfo{author}{\bibfnamefont{M.}~\bibnamefont{Bishof}},
  \bibinfo{author}{\bibfnamefont{X.}~\bibnamefont{Zhang}},
  \bibinfo{author}{\bibfnamefont{W.}~\bibnamefont{Zhang}},
  \bibinfo{author}{\bibfnamefont{S.}~\bibnamefont{Bromley}}, \bibnamefont{and}
  \bibinfo{author}{\bibfnamefont{J.}~\bibnamefont{Ye}},
  \bibinfo{journal}{Nature}  (\bibinfo{year}{2014}).

\bibitem[{\citenamefont{Calarco et~al.}(2000)\citenamefont{Calarco, Hinds,
  Jaksch, Schmiedmayer, Cirac, and Zoller}}]{Calarco2000}
\bibinfo{author}{\bibfnamefont{T.}~\bibnamefont{Calarco}},
  \bibinfo{author}{\bibfnamefont{E.}~\bibnamefont{Hinds}},
  \bibinfo{author}{\bibfnamefont{D.}~\bibnamefont{Jaksch}},
  \bibinfo{author}{\bibfnamefont{J.}~\bibnamefont{Schmiedmayer}},
  \bibinfo{author}{\bibfnamefont{J.}~\bibnamefont{Cirac}}, \bibnamefont{and}
  \bibinfo{author}{\bibfnamefont{P.}~\bibnamefont{Zoller}},
  \bibinfo{journal}{Physical Review A} \textbf{\bibinfo{volume}{61}},
  \bibinfo{pages}{022304} (\bibinfo{year}{2000}).

\bibitem[{\citenamefont{Kielpinski et~al.}(2002)\citenamefont{Kielpinski,
  Monroe, and Wineland}}]{Kielpinski2002}
\bibinfo{author}{\bibfnamefont{D.}~\bibnamefont{Kielpinski}},
  \bibinfo{author}{\bibfnamefont{C.}~\bibnamefont{Monroe}}, \bibnamefont{and}
  \bibinfo{author}{\bibfnamefont{D.~J.} \bibnamefont{Wineland}},
  \bibinfo{journal}{Nature} \textbf{\bibinfo{volume}{417}},
  \bibinfo{pages}{709} (\bibinfo{year}{2002}).

\bibitem[{\citenamefont{Bloch et~al.}(2012)\citenamefont{Bloch, Dalibard, and
  Nascimb{\`e}ne}}]{Bloch12}
\bibinfo{author}{\bibfnamefont{I.}~\bibnamefont{Bloch}},
  \bibinfo{author}{\bibfnamefont{J.}~\bibnamefont{Dalibard}}, \bibnamefont{and}
  \bibinfo{author}{\bibfnamefont{S.}~\bibnamefont{Nascimb{\`e}ne}},
  \bibinfo{journal}{Nature Physics} \textbf{\bibinfo{volume}{8}},
  \bibinfo{pages}{267} (\bibinfo{year}{2012}).

\bibitem[{\citenamefont{Peirce et~al.}(1988{\natexlab{a}})\citenamefont{Peirce,
  Dahleh, and Rabitz}}]{Peirce1988}
\bibinfo{author}{\bibfnamefont{A.~P.} \bibnamefont{Peirce}},
  \bibinfo{author}{\bibfnamefont{M.~A.} \bibnamefont{Dahleh}},
  \bibnamefont{and} \bibinfo{author}{\bibfnamefont{H.}~\bibnamefont{Rabitz}},
  \bibinfo{journal}{Physical Review A} \textbf{\bibinfo{volume}{37}},
  \bibinfo{pages}{4950} (\bibinfo{year}{1988}{\natexlab{a}}).

\bibitem[{\citenamefont{Koch et~al.}(2004)\citenamefont{Koch, Palao, Kosloff,
  and Masnou-Seeuws}}]{Koch2004}
\bibinfo{author}{\bibfnamefont{C.~P.} \bibnamefont{Koch}},
  \bibinfo{author}{\bibfnamefont{J.~P.} \bibnamefont{Palao}},
  \bibinfo{author}{\bibfnamefont{R.}~\bibnamefont{Kosloff}}, \bibnamefont{and}
  \bibinfo{author}{\bibfnamefont{F.}~\bibnamefont{Masnou-Seeuws}},
  \bibinfo{journal}{Physical Review A} \textbf{\bibinfo{volume}{70}},
  \bibinfo{pages}{013402} (\bibinfo{year}{2004}).

\bibitem[{\citenamefont{Rabitz et~al.}(2000)\citenamefont{Rabitz,
  de~Vivie-Riedle, Motzkus, and Kompa}}]{Rabitz2000}
\bibinfo{author}{\bibfnamefont{H.}~\bibnamefont{Rabitz}},
  \bibinfo{author}{\bibfnamefont{R.}~\bibnamefont{de~Vivie-Riedle}},
  \bibinfo{author}{\bibfnamefont{M.}~\bibnamefont{Motzkus}}, \bibnamefont{and}
  \bibinfo{author}{\bibfnamefont{K.}~\bibnamefont{Kompa}},
  \bibinfo{journal}{Science} \textbf{\bibinfo{volume}{288}},
  \bibinfo{pages}{824} (\bibinfo{year}{2000}).

\bibitem[{\citenamefont{Borzi et~al.}(2002)\citenamefont{Borzi, Stadler, and
  Hohenester}}]{Borzi2002}
\bibinfo{author}{\bibfnamefont{A.}~\bibnamefont{Borzi}},
  \bibinfo{author}{\bibfnamefont{G.}~\bibnamefont{Stadler}}, \bibnamefont{and}
  \bibinfo{author}{\bibfnamefont{U.}~\bibnamefont{Hohenester}},
  \bibinfo{journal}{Physical Review A} \textbf{\bibinfo{volume}{66}},
  \bibinfo{pages}{053811} (\bibinfo{year}{2002}).

\bibitem[{\citenamefont{Hohenester}(2006)}]{Hohenester2006}
\bibinfo{author}{\bibfnamefont{U.}~\bibnamefont{Hohenester}},
  \bibinfo{journal}{Physical Review B} \textbf{\bibinfo{volume}{74}},
  \bibinfo{pages}{161307} (\bibinfo{year}{2006}).

\bibitem[{\citenamefont{N{\"o}bauer et~al.}(2014)\citenamefont{N{\"o}bauer,
  Angerer, Bartels, Trupke, Rotter, Schmiedmayer, Mintert, and
  Majer}}]{Nobauer2014}
\bibinfo{author}{\bibfnamefont{T.}~\bibnamefont{N{\"o}bauer}},
  \bibinfo{author}{\bibfnamefont{A.}~\bibnamefont{Angerer}},
  \bibinfo{author}{\bibfnamefont{B.}~\bibnamefont{Bartels}},
  \bibinfo{author}{\bibfnamefont{M.}~\bibnamefont{Trupke}},
  \bibinfo{author}{\bibfnamefont{S.}~\bibnamefont{Rotter}},
  \bibinfo{author}{\bibfnamefont{J.}~\bibnamefont{Schmiedmayer}},
  \bibinfo{author}{\bibfnamefont{F.}~\bibnamefont{Mintert}}, \bibnamefont{and}
  \bibinfo{author}{\bibfnamefont{J.}~\bibnamefont{Majer}},
  \bibinfo{journal}{arXiv preprint arXiv:1412.5051}  (\bibinfo{year}{2014}).

\bibitem[{\citenamefont{Pethick and Smith}(2002)}]{Pethick2002}
\bibinfo{author}{\bibfnamefont{C.}~\bibnamefont{Pethick}} \bibnamefont{and}
  \bibinfo{author}{\bibfnamefont{H.}~\bibnamefont{Smith}},
  \emph{\bibinfo{title}{Bose-Einstein condensation in dilute gases}}
  (\bibinfo{publisher}{Cambridge university press}, \bibinfo{year}{2002}).

\bibitem[{\citenamefont{Dalfovo et~al.}(1999)\citenamefont{Dalfovo, Giorgini,
  Pitaevskii, and Stringari}}]{Dalfovo99}
\bibinfo{author}{\bibfnamefont{F.}~\bibnamefont{Dalfovo}},
  \bibinfo{author}{\bibfnamefont{S.}~\bibnamefont{Giorgini}},
  \bibinfo{author}{\bibfnamefont{L.~P.} \bibnamefont{Pitaevskii}},
  \bibnamefont{and}
  \bibinfo{author}{\bibfnamefont{S.}~\bibnamefont{Stringari}},
  \bibinfo{journal}{Reviews of Modern Physics} \textbf{\bibinfo{volume}{71}},
  \bibinfo{pages}{463} (\bibinfo{year}{1999}).

\bibitem[{\citenamefont{Hohenester et~al.}(2007)\citenamefont{Hohenester,
  Rekdal, Borz\`\i, and Schmiedmayer}}]{HoReBoSc07}
\bibinfo{author}{\bibfnamefont{U.}~\bibnamefont{Hohenester}},
  \bibinfo{author}{\bibfnamefont{P.~K.} \bibnamefont{Rekdal}},
  \bibinfo{author}{\bibfnamefont{A.}~\bibnamefont{Borz\`\i}}, \bibnamefont{and}
  \bibinfo{author}{\bibfnamefont{J.}~\bibnamefont{Schmiedmayer}},
  \bibinfo{journal}{Phys. Rev. A} \textbf{\bibinfo{volume}{75}},
  \bibinfo{pages}{023602} (\bibinfo{year}{2007}).

\bibitem[{\citenamefont{Peirce et~al.}(1988{\natexlab{b}})\citenamefont{Peirce,
  Dahleh, and Rabitz}}]{Peirce88}
\bibinfo{author}{\bibfnamefont{A.~P.} \bibnamefont{Peirce}},
  \bibinfo{author}{\bibfnamefont{M.~A.} \bibnamefont{Dahleh}},
  \bibnamefont{and} \bibinfo{author}{\bibfnamefont{H.}~\bibnamefont{Rabitz}},
  \bibinfo{journal}{Phys. Rev. A} \textbf{\bibinfo{volume}{37}},
  \bibinfo{pages}{4950} (\bibinfo{year}{1988}{\natexlab{b}}).

\bibitem[{\citenamefont{J\"ager and
  Hohenester}(2013{\natexlab{a}})}]{Jager2013}
\bibinfo{author}{\bibfnamefont{G.}~\bibnamefont{J\"ager}} \bibnamefont{and}
  \bibinfo{author}{\bibfnamefont{U.}~\bibnamefont{Hohenester}},
  \bibinfo{journal}{Phys. Rev. A} \textbf{\bibinfo{volume}{88}},
  \bibinfo{pages}{035601} (\bibinfo{year}{2013}{\natexlab{a}}).

\bibitem[{\citenamefont{B{\"u}cker et~al.}(2011)\citenamefont{B{\"u}cker,
  Grond, Manz, Berrada, Betz, Koller, Hohenester, Schumm, Perrin, and
  Schmiedmayer}}]{Bucker2011}
\bibinfo{author}{\bibfnamefont{R.}~\bibnamefont{B{\"u}cker}},
  \bibinfo{author}{\bibfnamefont{J.}~\bibnamefont{Grond}},
  \bibinfo{author}{\bibfnamefont{S.}~\bibnamefont{Manz}},
  \bibinfo{author}{\bibfnamefont{T.}~\bibnamefont{Berrada}},
  \bibinfo{author}{\bibfnamefont{T.}~\bibnamefont{Betz}},
  \bibinfo{author}{\bibfnamefont{C.}~\bibnamefont{Koller}},
  \bibinfo{author}{\bibfnamefont{U.}~\bibnamefont{Hohenester}},
  \bibinfo{author}{\bibfnamefont{T.}~\bibnamefont{Schumm}},
  \bibinfo{author}{\bibfnamefont{A.}~\bibnamefont{Perrin}}, \bibnamefont{and}
  \bibinfo{author}{\bibfnamefont{J.}~\bibnamefont{Schmiedmayer}},
  \bibinfo{journal}{Nature Physics} \textbf{\bibinfo{volume}{7}},
  \bibinfo{pages}{608} (\bibinfo{year}{2011}).

\bibitem[{\citenamefont{van Frank et~al.}(2014)\citenamefont{van Frank,
  Negretti, Berrada, B{\"u}cker, Montangero, Schaff, Schumm, Calarco, and
  Schmiedmayer}}]{vanFrank2014}
\bibinfo{author}{\bibfnamefont{S.}~\bibnamefont{van Frank}},
  \bibinfo{author}{\bibfnamefont{A.}~\bibnamefont{Negretti}},
  \bibinfo{author}{\bibfnamefont{T.}~\bibnamefont{Berrada}},
  \bibinfo{author}{\bibfnamefont{R.}~\bibnamefont{B{\"u}cker}},
  \bibinfo{author}{\bibfnamefont{S.}~\bibnamefont{Montangero}},
  \bibinfo{author}{\bibfnamefont{J.-F.} \bibnamefont{Schaff}},
  \bibinfo{author}{\bibfnamefont{T.}~\bibnamefont{Schumm}},
  \bibinfo{author}{\bibfnamefont{T.}~\bibnamefont{Calarco}}, \bibnamefont{and}
  \bibinfo{author}{\bibfnamefont{J.}~\bibnamefont{Schmiedmayer}},
  \bibinfo{journal}{Nature communications} \textbf{\bibinfo{volume}{5}}
  (\bibinfo{year}{2014}).

\bibitem[{\citenamefont{Hohenester}(2014)}]{Ho2014}
\bibinfo{author}{\bibfnamefont{U.}~\bibnamefont{Hohenester}},
  \bibinfo{journal}{Computer Physics Communications}
  \textbf{\bibinfo{volume}{185}}, \bibinfo{pages}{194 } (\bibinfo{year}{2014}),
  ISSN \bibinfo{issn}{0010-4655}.

\bibitem[{\citenamefont{von Winckel and Borzi}(2008)}]{WiBo08}
\bibinfo{author}{\bibfnamefont{G.}~\bibnamefont{von Winckel}} \bibnamefont{and}
  \bibinfo{author}{\bibfnamefont{A.}~\bibnamefont{Borzi}},
  \bibinfo{journal}{Inverse Problems} \textbf{\bibinfo{volume}{24}},
  \bibinfo{pages}{034007} (\bibinfo{year}{2008}).

\bibitem[{\citenamefont{B\"ucker et~al.}(2013)\citenamefont{B\"ucker, Berrada,
  van Frank, Schaff, Schumm, Schmiedmayer, J\"ager, Grond, and
  Hohenester}}]{BueBeFr13}
\bibinfo{author}{\bibfnamefont{R.}~\bibnamefont{B\"ucker}},
  \bibinfo{author}{\bibfnamefont{T.}~\bibnamefont{Berrada}},
  \bibinfo{author}{\bibfnamefont{S.}~\bibnamefont{van Frank}},
  \bibinfo{author}{\bibfnamefont{J.-F.} \bibnamefont{Schaff}},
  \bibinfo{author}{\bibfnamefont{T.}~\bibnamefont{Schumm}},
  \bibinfo{author}{\bibfnamefont{J.}~\bibnamefont{Schmiedmayer}},
  \bibinfo{author}{\bibfnamefont{G.}~\bibnamefont{J\"ager}},
  \bibinfo{author}{\bibfnamefont{J.}~\bibnamefont{Grond}}, \bibnamefont{and}
  \bibinfo{author}{\bibfnamefont{U.}~\bibnamefont{Hohenester}},
  \bibinfo{journal}{Journal of Physics B: Atomic, Molecular and Optical
  Physics} \textbf{\bibinfo{volume}{46}}, \bibinfo{pages}{104012}
  (\bibinfo{year}{2013}).

\bibitem[{\citenamefont{J\"ager and
  Hohenester}(2013{\natexlab{b}})}]{JaeHo2013}
\bibinfo{author}{\bibfnamefont{G.}~\bibnamefont{J\"ager}} \bibnamefont{and}
  \bibinfo{author}{\bibfnamefont{U.}~\bibnamefont{Hohenester}},
  \bibinfo{journal}{Phys. Rev. A} \textbf{\bibinfo{volume}{88}},
  \bibinfo{pages}{035601} (\bibinfo{year}{2013}{\natexlab{b}}).

\bibitem[{\citenamefont{J\"ager et~al.}(2014)\citenamefont{J\"ager, Reich,
  Goerz, Koch, and Hohenester}}]{JaeReGoKoHo2014}
\bibinfo{author}{\bibfnamefont{G.}~\bibnamefont{J\"ager}},
  \bibinfo{author}{\bibfnamefont{D.~M.} \bibnamefont{Reich}},
  \bibinfo{author}{\bibfnamefont{M.~H.} \bibnamefont{Goerz}},
  \bibinfo{author}{\bibfnamefont{C.~P.} \bibnamefont{Koch}}, \bibnamefont{and}
  \bibinfo{author}{\bibfnamefont{U.}~\bibnamefont{Hohenester}},
  \bibinfo{journal}{Phys. Rev. A} \textbf{\bibinfo{volume}{90}},
  \bibinfo{pages}{033628} (\bibinfo{year}{2014}).

\bibitem[{\citenamefont{Hager and Zhang}(2005)}]{HaZh2005}
\bibinfo{author}{\bibfnamefont{W.}~\bibnamefont{Hager}} \bibnamefont{and}
  \bibinfo{author}{\bibfnamefont{H.}~\bibnamefont{Zhang}},
  \bibinfo{journal}{SIAM Journal on Optimization}
  \textbf{\bibinfo{volume}{16}}, \bibinfo{pages}{170} (\bibinfo{year}{2005}).

\bibitem[{\citenamefont{Bao et~al.}(2003)\citenamefont{Bao, Jaksch, and
  Markowich}}]{BaJaMa2003}
\bibinfo{author}{\bibfnamefont{W.}~\bibnamefont{Bao}},
  \bibinfo{author}{\bibfnamefont{D.}~\bibnamefont{Jaksch}}, \bibnamefont{and}
  \bibinfo{author}{\bibfnamefont{P.~A.} \bibnamefont{Markowich}},
  \bibinfo{journal}{Journal of Computational Physics}
  \textbf{\bibinfo{volume}{187}}, \bibinfo{pages}{318 } (\bibinfo{year}{2003}).

\bibitem[{vid()}]{videos}
\bibinfo{note}{A three-dimensional visualization of the numerical examples
  considered in this article is available online},
  \urlprefix\url{https://www.youtube.com/watch?v=DEcnqSwPgrw}.

\bibitem[{\citenamefont{Ketterle et~al.}(1999)\citenamefont{Ketterle, Durfee,
  and Stamper-Kurn}}]{Ketterle1999}
\bibinfo{author}{\bibfnamefont{W.}~\bibnamefont{Ketterle}},
  \bibinfo{author}{\bibfnamefont{D.}~\bibnamefont{Durfee}}, \bibnamefont{and}
  \bibinfo{author}{\bibfnamefont{D.}~\bibnamefont{Stamper-Kurn}},
  \bibinfo{journal}{arXiv preprint cond-mat/9904034}
  \textbf{\bibinfo{volume}{5}} (\bibinfo{year}{1999}).

\bibitem[{\citenamefont{Schaff et~al.}(2011)\citenamefont{Schaff, Song,
  Capuzzi, Vignolo, and Labeyrie}}]{Schaff2011}
\bibinfo{author}{\bibfnamefont{J.-F.} \bibnamefont{Schaff}},
  \bibinfo{author}{\bibfnamefont{X.-L.} \bibnamefont{Song}},
  \bibinfo{author}{\bibfnamefont{P.}~\bibnamefont{Capuzzi}},
  \bibinfo{author}{\bibfnamefont{P.}~\bibnamefont{Vignolo}}, \bibnamefont{and}
  \bibinfo{author}{\bibfnamefont{G.}~\bibnamefont{Labeyrie}},
  \bibinfo{journal}{EPL (Europhysics Letters)} \textbf{\bibinfo{volume}{93}},
  \bibinfo{pages}{23001} (\bibinfo{year}{2011}).

\bibitem[{\citenamefont{Seaman et~al.}(2007)\citenamefont{Seaman, Kr\"amer,
  Anderson, and Holland}}]{Seaman2007}
\bibinfo{author}{\bibfnamefont{B.~T.} \bibnamefont{Seaman}},
  \bibinfo{author}{\bibfnamefont{M.}~\bibnamefont{Kr\"amer}},
  \bibinfo{author}{\bibfnamefont{D.~Z.} \bibnamefont{Anderson}},
  \bibnamefont{and} \bibinfo{author}{\bibfnamefont{M.~J.}
  \bibnamefont{Holland}}, \bibinfo{journal}{Phys. Rev. A}
  \textbf{\bibinfo{volume}{75}}, \bibinfo{pages}{023615}
  (\bibinfo{year}{2007}).

\bibitem[{\citenamefont{Henderson et~al.}(2009)\citenamefont{Henderson, Ryu,
  MacCormick, and Boshier}}]{Henderson2009}
\bibinfo{author}{\bibfnamefont{K.}~\bibnamefont{Henderson}},
  \bibinfo{author}{\bibfnamefont{C.}~\bibnamefont{Ryu}},
  \bibinfo{author}{\bibfnamefont{C.}~\bibnamefont{MacCormick}},
  \bibnamefont{and} \bibinfo{author}{\bibfnamefont{M.}~\bibnamefont{Boshier}},
  \bibinfo{journal}{New Journal of Physics} \textbf{\bibinfo{volume}{11}},
  \bibinfo{pages}{043030} (\bibinfo{year}{2009}).

\bibitem[{\citenamefont{Shvarchuck et~al.}(2002)\citenamefont{Shvarchuck,
  Buggle, Petrov, Dieckmann, Zielonkowski, Kemmann, Tiecke, von Klitzing,
  Shlyapnikov, and Walraven}}]{Shvarchuck2002}
\bibinfo{author}{\bibfnamefont{I.}~\bibnamefont{Shvarchuck}},
  \bibinfo{author}{\bibfnamefont{C.}~\bibnamefont{Buggle}},
  \bibinfo{author}{\bibfnamefont{D.~S.} \bibnamefont{Petrov}},
  \bibinfo{author}{\bibfnamefont{K.}~\bibnamefont{Dieckmann}},
  \bibinfo{author}{\bibfnamefont{M.}~\bibnamefont{Zielonkowski}},
  \bibinfo{author}{\bibfnamefont{M.}~\bibnamefont{Kemmann}},
  \bibinfo{author}{\bibfnamefont{T.~G.} \bibnamefont{Tiecke}},
  \bibinfo{author}{\bibfnamefont{W.}~\bibnamefont{von Klitzing}},
  \bibinfo{author}{\bibfnamefont{G.~V.} \bibnamefont{Shlyapnikov}},
  \bibnamefont{and} \bibinfo{author}{\bibfnamefont{J.~T.~M.}
  \bibnamefont{Walraven}}, \bibinfo{journal}{Phys. Rev. Lett.}
  \textbf{\bibinfo{volume}{89}}, \bibinfo{pages}{270404}
  (\bibinfo{year}{2002}).

\bibitem[{\citenamefont{Ryu et~al.}(2007{\natexlab{a}})\citenamefont{Ryu,
  Andersen, Clad{\'e}, Natarajan, Helmerson, and Phillips}}]{Ryu2007}
\bibinfo{author}{\bibfnamefont{C.}~\bibnamefont{Ryu}},
  \bibinfo{author}{\bibfnamefont{M.}~\bibnamefont{Andersen}},
  \bibinfo{author}{\bibfnamefont{P.}~\bibnamefont{Clad{\'e}}},
  \bibinfo{author}{\bibfnamefont{V.}~\bibnamefont{Natarajan}},
  \bibinfo{author}{\bibfnamefont{K.}~\bibnamefont{Helmerson}},
  \bibnamefont{and} \bibinfo{author}{\bibfnamefont{W.}~\bibnamefont{Phillips}},
  \bibinfo{journal}{Physical Review Letters} \textbf{\bibinfo{volume}{99}},
  \bibinfo{pages}{260401} (\bibinfo{year}{2007}{\natexlab{a}}).

\bibitem[{\citenamefont{Ryu et~al.}(2013)\citenamefont{Ryu, Blackburn, Blinova,
  and Boshier}}]{Ryu2013}
\bibinfo{author}{\bibfnamefont{C.}~\bibnamefont{Ryu}},
  \bibinfo{author}{\bibfnamefont{P.~W.} \bibnamefont{Blackburn}},
  \bibinfo{author}{\bibfnamefont{A.~A.} \bibnamefont{Blinova}},
  \bibnamefont{and} \bibinfo{author}{\bibfnamefont{M.~G.}
  \bibnamefont{Boshier}}, \bibinfo{journal}{Phys. Rev. Lett.}
  \textbf{\bibinfo{volume}{111}}, \bibinfo{pages}{205301}
  (\bibinfo{year}{2013}).

\bibitem[{\citenamefont{Beattie et~al.}(2013)\citenamefont{Beattie, Moulder,
  Fletcher, and Hadzibabic}}]{Beattie2013}
\bibinfo{author}{\bibfnamefont{S.}~\bibnamefont{Beattie}},
  \bibinfo{author}{\bibfnamefont{S.}~\bibnamefont{Moulder}},
  \bibinfo{author}{\bibfnamefont{R.~J.} \bibnamefont{Fletcher}},
  \bibnamefont{and}
  \bibinfo{author}{\bibfnamefont{Z.}~\bibnamefont{Hadzibabic}},
  \bibinfo{journal}{Phys. Rev. Lett.} \textbf{\bibinfo{volume}{110}},
  \bibinfo{pages}{025301} (\bibinfo{year}{2013}).

\bibitem[{\citenamefont{Jendrzejewski et~al.}(2014)\citenamefont{Jendrzejewski,
  Eckel, Murray, Lanier, Edwards, Lobb, and Campbell}}]{Jendrzejewski2014}
\bibinfo{author}{\bibfnamefont{F.}~\bibnamefont{Jendrzejewski}},
  \bibinfo{author}{\bibfnamefont{S.}~\bibnamefont{Eckel}},
  \bibinfo{author}{\bibfnamefont{N.}~\bibnamefont{Murray}},
  \bibinfo{author}{\bibfnamefont{C.}~\bibnamefont{Lanier}},
  \bibinfo{author}{\bibfnamefont{M.}~\bibnamefont{Edwards}},
  \bibinfo{author}{\bibfnamefont{C.~J.} \bibnamefont{Lobb}}, \bibnamefont{and}
  \bibinfo{author}{\bibfnamefont{G.~K.} \bibnamefont{Campbell}},
  \bibinfo{journal}{Phys. Rev. Lett.} \textbf{\bibinfo{volume}{113}},
  \bibinfo{pages}{045305} (\bibinfo{year}{2014}).

\bibitem[{\citenamefont{Eckel et~al.}(2014)\citenamefont{Eckel, Lee,
  Jendrzejewski, Murray, Clark, Lobb, Phillips, Edwards, and
  Campbell}}]{Eckel2014}
\bibinfo{author}{\bibfnamefont{S.}~\bibnamefont{Eckel}},
  \bibinfo{author}{\bibfnamefont{J.~G.} \bibnamefont{Lee}},
  \bibinfo{author}{\bibfnamefont{F.}~\bibnamefont{Jendrzejewski}},
  \bibinfo{author}{\bibfnamefont{N.}~\bibnamefont{Murray}},
  \bibinfo{author}{\bibfnamefont{C.~W.} \bibnamefont{Clark}},
  \bibinfo{author}{\bibfnamefont{C.~J.} \bibnamefont{Lobb}},
  \bibinfo{author}{\bibfnamefont{W.~D.} \bibnamefont{Phillips}},
  \bibinfo{author}{\bibfnamefont{M.}~\bibnamefont{Edwards}}, \bibnamefont{and}
  \bibinfo{author}{\bibfnamefont{G.~K.} \bibnamefont{Campbell}},
  \bibinfo{journal}{Nature} \textbf{\bibinfo{volume}{506}},
  \bibinfo{pages}{200} (\bibinfo{year}{2014}).

\bibitem[{\citenamefont{Ryu et~al.}(2007{\natexlab{b}})\citenamefont{Ryu,
  Andersen, Clad\'e, Natarajan, Helmerson, and Phillips}}]{RyuAndCla2007}
\bibinfo{author}{\bibfnamefont{C.}~\bibnamefont{Ryu}},
  \bibinfo{author}{\bibfnamefont{M.~F.} \bibnamefont{Andersen}},
  \bibinfo{author}{\bibfnamefont{P.}~\bibnamefont{Clad\'e}},
  \bibinfo{author}{\bibfnamefont{V.}~\bibnamefont{Natarajan}},
  \bibinfo{author}{\bibfnamefont{K.}~\bibnamefont{Helmerson}},
  \bibnamefont{and} \bibinfo{author}{\bibfnamefont{W.~D.}
  \bibnamefont{Phillips}}, \bibinfo{journal}{Phys. Rev. Lett.}
  \textbf{\bibinfo{volume}{99}}, \bibinfo{pages}{260401}
  (\bibinfo{year}{2007}{\natexlab{b}}).

\bibitem[{\citenamefont{Lesanovsky
  et~al.}(2006{\natexlab{a}})\citenamefont{Lesanovsky, Schumm, Hofferberth,
  Andersson, Kr\"uger, and Schmiedmayer}}]{LesSchHoff2006}
\bibinfo{author}{\bibfnamefont{I.}~\bibnamefont{Lesanovsky}},
  \bibinfo{author}{\bibfnamefont{T.}~\bibnamefont{Schumm}},
  \bibinfo{author}{\bibfnamefont{S.}~\bibnamefont{Hofferberth}},
  \bibinfo{author}{\bibfnamefont{L.~M.} \bibnamefont{Andersson}},
  \bibinfo{author}{\bibfnamefont{P.}~\bibnamefont{Kr\"uger}}, \bibnamefont{and}
  \bibinfo{author}{\bibfnamefont{J.}~\bibnamefont{Schmiedmayer}},
  \bibinfo{journal}{Phys. Rev. A} \textbf{\bibinfo{volume}{73}},
  \bibinfo{pages}{033619} (\bibinfo{year}{2006}{\natexlab{a}}).

\bibitem[{\citenamefont{Schumm et~al.}(2005)\citenamefont{Schumm, Hofferberth,
  Andersson, Wildermuth, Groth, Bar-Joseph, Schmiedmayer, and
  Kr{\"u}ger}}]{Schumm2005}
\bibinfo{author}{\bibfnamefont{T.}~\bibnamefont{Schumm}},
  \bibinfo{author}{\bibfnamefont{S.}~\bibnamefont{Hofferberth}},
  \bibinfo{author}{\bibfnamefont{L.~M.} \bibnamefont{Andersson}},
  \bibinfo{author}{\bibfnamefont{S.}~\bibnamefont{Wildermuth}},
  \bibinfo{author}{\bibfnamefont{S.}~\bibnamefont{Groth}},
  \bibinfo{author}{\bibfnamefont{I.}~\bibnamefont{Bar-Joseph}},
  \bibinfo{author}{\bibfnamefont{J.}~\bibnamefont{Schmiedmayer}},
  \bibnamefont{and}
  \bibinfo{author}{\bibfnamefont{P.}~\bibnamefont{Kr{\"u}ger}},
  \bibinfo{journal}{Nature Physics} \textbf{\bibinfo{volume}{1}},
  \bibinfo{pages}{57} (\bibinfo{year}{2005}).

\bibitem[{\citenamefont{Folman et~al.}(2002)\citenamefont{Folman, Kr{\"u}ger,
  Schmiedmayer, Denschlag, and Henkel}}]{Folman2002}
\bibinfo{author}{\bibfnamefont{R.}~\bibnamefont{Folman}},
  \bibinfo{author}{\bibfnamefont{P.}~\bibnamefont{Kr{\"u}ger}},
  \bibinfo{author}{\bibfnamefont{J.}~\bibnamefont{Schmiedmayer}},
  \bibinfo{author}{\bibfnamefont{J.}~\bibnamefont{Denschlag}},
  \bibnamefont{and} \bibinfo{author}{\bibfnamefont{C.}~\bibnamefont{Henkel}},
  \bibinfo{journal}{Advances in Atomic, Molecular, and Optical Physics}
  \textbf{\bibinfo{volume}{48}}, \bibinfo{pages}{263} (\bibinfo{year}{2002}).

\bibitem[{\citenamefont{Reichel and Vuletic}(2010)}]{Reichel2010}
\bibinfo{author}{\bibfnamefont{J.}~\bibnamefont{Reichel}} \bibnamefont{and}
  \bibinfo{author}{\bibfnamefont{V.}~\bibnamefont{Vuletic}},
  \emph{\bibinfo{title}{Atom Chips}} (\bibinfo{publisher}{John Wiley \& Sons},
  \bibinfo{year}{2010}).

\bibitem[{\citenamefont{Gring et~al.}(2012)\citenamefont{Gring, Kuhnert,
  Langen, Kitagawa, Rauer, Schreitl, Mazets, Smith, Demler, and
  Schmiedmayer}}]{Gring2012}
\bibinfo{author}{\bibfnamefont{M.}~\bibnamefont{Gring}},
  \bibinfo{author}{\bibfnamefont{M.}~\bibnamefont{Kuhnert}},
  \bibinfo{author}{\bibfnamefont{T.}~\bibnamefont{Langen}},
  \bibinfo{author}{\bibfnamefont{T.}~\bibnamefont{Kitagawa}},
  \bibinfo{author}{\bibfnamefont{B.}~\bibnamefont{Rauer}},
  \bibinfo{author}{\bibfnamefont{M.}~\bibnamefont{Schreitl}},
  \bibinfo{author}{\bibfnamefont{I.}~\bibnamefont{Mazets}},
  \bibinfo{author}{\bibfnamefont{D.~A.} \bibnamefont{Smith}},
  \bibinfo{author}{\bibfnamefont{E.}~\bibnamefont{Demler}}, \bibnamefont{and}
  \bibinfo{author}{\bibfnamefont{J.}~\bibnamefont{Schmiedmayer}},
  \bibinfo{journal}{Science} \textbf{\bibinfo{volume}{337}},
  \bibinfo{pages}{1318} (\bibinfo{year}{2012}).

\bibitem[{\citenamefont{Kuhnert et~al.}(2013)\citenamefont{Kuhnert, Geiger,
  Langen, Gring, Rauer, Kitagawa, Demler, Adu~Smith, and
  Schmiedmayer}}]{Kuhnert2013}
\bibinfo{author}{\bibfnamefont{M.}~\bibnamefont{Kuhnert}},
  \bibinfo{author}{\bibfnamefont{R.}~\bibnamefont{Geiger}},
  \bibinfo{author}{\bibfnamefont{T.}~\bibnamefont{Langen}},
  \bibinfo{author}{\bibfnamefont{M.}~\bibnamefont{Gring}},
  \bibinfo{author}{\bibfnamefont{B.}~\bibnamefont{Rauer}},
  \bibinfo{author}{\bibfnamefont{T.}~\bibnamefont{Kitagawa}},
  \bibinfo{author}{\bibfnamefont{E.}~\bibnamefont{Demler}},
  \bibinfo{author}{\bibfnamefont{D.}~\bibnamefont{Adu~Smith}},
  \bibnamefont{and}
  \bibinfo{author}{\bibfnamefont{J.}~\bibnamefont{Schmiedmayer}},
  \bibinfo{journal}{Phys. Rev. Lett.} \textbf{\bibinfo{volume}{110}},
  \bibinfo{pages}{090405} (\bibinfo{year}{2013}).

\bibitem[{\citenamefont{Smith et~al.}(2013)\citenamefont{Smith, Gring, Langen,
  Kuhnert, Rauer, Geiger, Kitagawa, Mazets, Demler, and
  Schmiedmayer}}]{Smith2013}
\bibinfo{author}{\bibfnamefont{D.~A.} \bibnamefont{Smith}},
  \bibinfo{author}{\bibfnamefont{M.}~\bibnamefont{Gring}},
  \bibinfo{author}{\bibfnamefont{T.}~\bibnamefont{Langen}},
  \bibinfo{author}{\bibfnamefont{M.}~\bibnamefont{Kuhnert}},
  \bibinfo{author}{\bibfnamefont{B.}~\bibnamefont{Rauer}},
  \bibinfo{author}{\bibfnamefont{R.}~\bibnamefont{Geiger}},
  \bibinfo{author}{\bibfnamefont{T.}~\bibnamefont{Kitagawa}},
  \bibinfo{author}{\bibfnamefont{I.}~\bibnamefont{Mazets}},
  \bibinfo{author}{\bibfnamefont{E.}~\bibnamefont{Demler}}, \bibnamefont{and}
  \bibinfo{author}{\bibfnamefont{J.}~\bibnamefont{Schmiedmayer}},
  \bibinfo{journal}{New Journal of Physics} \textbf{\bibinfo{volume}{15}},
  \bibinfo{pages}{075011} (\bibinfo{year}{2013}).

\bibitem[{\citenamefont{Langen et~al.}(2013{\natexlab{a}})\citenamefont{Langen,
  Gring, Kuhnert, Rauer, Geiger, Smith, Mazets, and
  Schmiedmayer}}]{Langen2013b}
\bibinfo{author}{\bibfnamefont{T.}~\bibnamefont{Langen}},
  \bibinfo{author}{\bibfnamefont{M.}~\bibnamefont{Gring}},
  \bibinfo{author}{\bibfnamefont{M.}~\bibnamefont{Kuhnert}},
  \bibinfo{author}{\bibfnamefont{B.}~\bibnamefont{Rauer}},
  \bibinfo{author}{\bibfnamefont{R.}~\bibnamefont{Geiger}},
  \bibinfo{author}{\bibfnamefont{D.~A.} \bibnamefont{Smith}},
  \bibinfo{author}{\bibfnamefont{I.~E.} \bibnamefont{Mazets}},
  \bibnamefont{and}
  \bibinfo{author}{\bibfnamefont{J.}~\bibnamefont{Schmiedmayer}},
  \bibinfo{journal}{The European Physical Journal Special Topics}
  \textbf{\bibinfo{volume}{217}}, \bibinfo{pages}{43}
  (\bibinfo{year}{2013}{\natexlab{a}}).

\bibitem[{\citenamefont{Langen et~al.}(2014)\citenamefont{Langen, Erne, Geiger,
  Rauer, Schweigler, Kuhnert, Rohringer, Mazets, Gasenzer, and
  Schmiedmayer}}]{Langen2015}
\bibinfo{author}{\bibfnamefont{T.}~\bibnamefont{Langen}},
  \bibinfo{author}{\bibfnamefont{S.}~\bibnamefont{Erne}},
  \bibinfo{author}{\bibfnamefont{R.}~\bibnamefont{Geiger}},
  \bibinfo{author}{\bibfnamefont{B.}~\bibnamefont{Rauer}},
  \bibinfo{author}{\bibfnamefont{T.}~\bibnamefont{Schweigler}},
  \bibinfo{author}{\bibfnamefont{M.}~\bibnamefont{Kuhnert}},
  \bibinfo{author}{\bibfnamefont{W.}~\bibnamefont{Rohringer}},
  \bibinfo{author}{\bibfnamefont{I.~E.} \bibnamefont{Mazets}},
  \bibinfo{author}{\bibfnamefont{T.}~\bibnamefont{Gasenzer}}, \bibnamefont{and}
  \bibinfo{author}{\bibfnamefont{J.}~\bibnamefont{Schmiedmayer}},
  \bibinfo{journal}{arXiv:1411.7185}  (\bibinfo{year}{2014}).

\bibitem[{\citenamefont{Langen et~al.}(2013{\natexlab{b}})\citenamefont{Langen,
  Geiger, Kuhnert, Rauer, and Schmiedmayer}}]{Langen2013}
\bibinfo{author}{\bibfnamefont{T.}~\bibnamefont{Langen}},
  \bibinfo{author}{\bibfnamefont{R.}~\bibnamefont{Geiger}},
  \bibinfo{author}{\bibfnamefont{M.}~\bibnamefont{Kuhnert}},
  \bibinfo{author}{\bibfnamefont{B.}~\bibnamefont{Rauer}}, \bibnamefont{and}
  \bibinfo{author}{\bibfnamefont{J.}~\bibnamefont{Schmiedmayer}},
  \bibinfo{journal}{Nature Physics}  (\bibinfo{year}{2013}{\natexlab{b}}).

\bibitem[{\citenamefont{Geiger et~al.}(2014)\citenamefont{Geiger, Langen,
  Mazets, and Schmiedmayer}}]{Geiger2014}
\bibinfo{author}{\bibfnamefont{R.}~\bibnamefont{Geiger}},
  \bibinfo{author}{\bibfnamefont{T.}~\bibnamefont{Langen}},
  \bibinfo{author}{\bibfnamefont{I.}~\bibnamefont{Mazets}}, \bibnamefont{and}
  \bibinfo{author}{\bibfnamefont{J.}~\bibnamefont{Schmiedmayer}},
  \bibinfo{journal}{New Journal of Physics} \textbf{\bibinfo{volume}{16}},
  \bibinfo{pages}{053034} (\bibinfo{year}{2014}).

\bibitem[{\citenamefont{Berrada et~al.}(2013)\citenamefont{Berrada, van Frank,
  B{\"u}cker, Schumm, Schaff, and Schmiedmayer}}]{Berrada2013}
\bibinfo{author}{\bibfnamefont{T.}~\bibnamefont{Berrada}},
  \bibinfo{author}{\bibfnamefont{S.}~\bibnamefont{van Frank}},
  \bibinfo{author}{\bibfnamefont{R.}~\bibnamefont{B{\"u}cker}},
  \bibinfo{author}{\bibfnamefont{T.}~\bibnamefont{Schumm}},
  \bibinfo{author}{\bibfnamefont{J.-F.} \bibnamefont{Schaff}},
  \bibnamefont{and}
  \bibinfo{author}{\bibfnamefont{J.}~\bibnamefont{Schmiedmayer}},
  \bibinfo{journal}{Nature communications} \textbf{\bibinfo{volume}{4}}
  (\bibinfo{year}{2013}).

\bibitem[{\citenamefont{Grond et~al.}(2009)\citenamefont{Grond, von Winckel,
  Schmiedmayer, and Hohenester}}]{Grond2009}
\bibinfo{author}{\bibfnamefont{J.}~\bibnamefont{Grond}},
  \bibinfo{author}{\bibfnamefont{G.}~\bibnamefont{von Winckel}},
  \bibinfo{author}{\bibfnamefont{J.}~\bibnamefont{Schmiedmayer}},
  \bibnamefont{and}
  \bibinfo{author}{\bibfnamefont{U.}~\bibnamefont{Hohenester}},
  \bibinfo{journal}{Physical Review A} \textbf{\bibinfo{volume}{80}},
  \bibinfo{pages}{053625} (\bibinfo{year}{2009}).

\bibitem[{\citenamefont{Lesanovsky
  et~al.}(2006{\natexlab{b}})\citenamefont{Lesanovsky, Schumm, Hofferberth,
  Andersson, Kr{\"u}ger, and Schmiedmayer}}]{Lesanovski2006}
\bibinfo{author}{\bibfnamefont{I.}~\bibnamefont{Lesanovsky}},
  \bibinfo{author}{\bibfnamefont{T.}~\bibnamefont{Schumm}},
  \bibinfo{author}{\bibfnamefont{S.}~\bibnamefont{Hofferberth}},
  \bibinfo{author}{\bibfnamefont{L.~M.} \bibnamefont{Andersson}},
  \bibinfo{author}{\bibfnamefont{P.}~\bibnamefont{Kr{\"u}ger}},
  \bibnamefont{and}
  \bibinfo{author}{\bibfnamefont{J.}~\bibnamefont{Schmiedmayer}},
  \bibinfo{journal}{Physical Review A} \textbf{\bibinfo{volume}{73}},
  \bibinfo{pages}{033619} (\bibinfo{year}{2006}{\natexlab{b}}).

\bibitem[{\citenamefont{Leggett}(2001)}]{Le2001}
\bibinfo{author}{\bibfnamefont{A.~J.} \bibnamefont{Leggett}},
  \bibinfo{journal}{Rev. Mod. Phys.} \textbf{\bibinfo{volume}{73}},
  \bibinfo{pages}{307} (\bibinfo{year}{2001}).

\bibitem[{\citenamefont{Takahashi et~al.}(2014)\citenamefont{Takahashi,
  Nakamura, and Yamanaka}}]{TaNaYa2014}
\bibinfo{author}{\bibfnamefont{J.}~\bibnamefont{Takahashi}},
  \bibinfo{author}{\bibfnamefont{Y.}~\bibnamefont{Nakamura}}, \bibnamefont{and}
  \bibinfo{author}{\bibfnamefont{Y.}~\bibnamefont{Yamanaka}},
  \bibinfo{journal}{Annals of Physics} \textbf{\bibinfo{volume}{347}},
  \bibinfo{pages}{250 } (\bibinfo{year}{2014}), ISSN \bibinfo{issn}{0003-4916}.

\bibitem[{\citenamefont{Huepe et~al.}(2003)\citenamefont{Huepe, Tuckerman,
  M\'etens, and Brachet}}]{HuTuMeBra2003}
\bibinfo{author}{\bibfnamefont{C.}~\bibnamefont{Huepe}},
  \bibinfo{author}{\bibfnamefont{L.}~\bibnamefont{Tuckerman}},
  \bibinfo{author}{\bibfnamefont{S.}~\bibnamefont{M\'etens}}, \bibnamefont{and}
  \bibinfo{author}{\bibfnamefont{M.}~\bibnamefont{Brachet}},
  \bibinfo{journal}{Phys. Rev. A} \textbf{\bibinfo{volume}{68}},
  \bibinfo{pages}{023609} (\bibinfo{year}{2003}).

\bibitem[{\citenamefont{Langen}(2013)}]{LangenThesis}
\bibinfo{author}{\bibfnamefont{T.}~\bibnamefont{Langen}}, Ph.D. thesis,
  \bibinfo{school}{Technische Universit{\"a}t Wien} (\bibinfo{year}{2013}).

\bibitem[{\citenamefont{Meyer et~al.}(1990)\citenamefont{Meyer, Manthe, and
  Cederbaum}}]{Meyer1990}
\bibinfo{author}{\bibfnamefont{H.-D.} \bibnamefont{Meyer}},
  \bibinfo{author}{\bibfnamefont{U.}~\bibnamefont{Manthe}}, \bibnamefont{and}
  \bibinfo{author}{\bibfnamefont{L.~S.} \bibnamefont{Cederbaum}},
  \bibinfo{journal}{Chemical Physics Letters} \textbf{\bibinfo{volume}{165}},
  \bibinfo{pages}{73} (\bibinfo{year}{1990}).

\bibitem[{\citenamefont{Squire and Trapp}(1998)}]{SquTra1998}
\bibinfo{author}{\bibfnamefont{W.}~\bibnamefont{Squire}} \bibnamefont{and}
  \bibinfo{author}{\bibfnamefont{G.}~\bibnamefont{Trapp}},
  \bibinfo{journal}{SIAM Review} \textbf{\bibinfo{volume}{40}},
  \bibinfo{pages}{110} (\bibinfo{year}{1998}).

\bibitem[{\citenamefont{Chiofalo et~al.}(2000)\citenamefont{Chiofalo, Succi,
  and Tosi}}]{ChiSucTos2000}
\bibinfo{author}{\bibfnamefont{M.~L.} \bibnamefont{Chiofalo}},
  \bibinfo{author}{\bibfnamefont{S.}~\bibnamefont{Succi}}, \bibnamefont{and}
  \bibinfo{author}{\bibfnamefont{M.~P.} \bibnamefont{Tosi}},
  \bibinfo{journal}{Phys. Rev. E} \textbf{\bibinfo{volume}{62}},
  \bibinfo{pages}{7438} (\bibinfo{year}{2000}).

\bibitem[{\citenamefont{Bao and Du}(2004)}]{BaDu2004}
\bibinfo{author}{\bibfnamefont{W.}~\bibnamefont{Bao}} \bibnamefont{and}
  \bibinfo{author}{\bibfnamefont{Q.}~\bibnamefont{Du}}, \bibinfo{journal}{SIAM
  Journal on Scientific Computing} \textbf{\bibinfo{volume}{25}},
  \bibinfo{pages}{1674} (\bibinfo{year}{2004}).

\bibitem[{\citenamefont{Ronen et~al.}(2006)\citenamefont{Ronen, Bortolotti, and
  Bohn}}]{RoBoBo2006}
\bibinfo{author}{\bibfnamefont{S.}~\bibnamefont{Ronen}},
  \bibinfo{author}{\bibfnamefont{D.}~\bibnamefont{Bortolotti}},
  \bibnamefont{and} \bibinfo{author}{\bibfnamefont{J.}~\bibnamefont{Bohn}},
  \bibinfo{journal}{Phys. Rev. A} \textbf{\bibinfo{volume}{74}},
  \bibinfo{pages}{013623} (\bibinfo{year}{2006}).

\bibitem[{\citenamefont{Salasnich et~al.}(2002)\citenamefont{Salasnich, Parola,
  and Reatto}}]{Salasnich2002}
\bibinfo{author}{\bibfnamefont{L.}~\bibnamefont{Salasnich}},
  \bibinfo{author}{\bibfnamefont{A.}~\bibnamefont{Parola}}, \bibnamefont{and}
  \bibinfo{author}{\bibfnamefont{L.}~\bibnamefont{Reatto}},
  \bibinfo{journal}{Physical Review A} \textbf{\bibinfo{volume}{65}},
  \bibinfo{pages}{043614} (\bibinfo{year}{2002}).

\bibitem[{\citenamefont{Olshanii}(1998)}]{Olshanii1998}
\bibinfo{author}{\bibfnamefont{M.}~\bibnamefont{Olshanii}},
  \bibinfo{journal}{Physical Review Letters} \textbf{\bibinfo{volume}{81}},
  \bibinfo{pages}{938} (\bibinfo{year}{1998}).

\end{thebibliography}

\newpage
\section*{APPENDIX}

\subsection{Gradient of the reduced cost functional}
\label{ap:gradient_H_0_1}

In the main text we have introduced the cost functional $J(\blambda,\psi)$ in~\eqref{eq:cost_functional}
and the reduced cost functional $\hat J(\blambda)=J(\blambda,\psi_\blambda)$ in~\eqref{eq:reduced_cost_functional}.
Recall that,
for a given control $\blambda:(0,T)\to\R^m$ satisfying the boundary conditions \eqref{eq:boundary_conditions_lambda},
$\psi_\blambda$ is the solution to the initial value problem~\eqref{eq:state_equation} and~\eqref{eq:initial_condition_state_equation}
for the GPE with the corresponding potential $V_\blambda$.
Below, we argue why the $H^1$-gradient of $\hat J$ is given by 
the component $\BLambda:(0,T)\to\R^m$ of the solution $(\psi,p,\BLambda)$ to the system 
consisting of \eqref{eq:state_equation},~\eqref{eq:adjoint_equation} and
\begin{equation}
  \label{eq:gradient_corrector}
  \frac{d^2}{dt^2} (\BLambda-\gamma\blambda) = \Re \langle p, (\partial_\blambda V_\blambda) \,\psi \rangle,
\end{equation}
subject to the initial and terminal conditions~\eqref{eq:initial_condition_state_equation},
\eqref{eq:terminal_condition_adjoint_equation} and
\begin{equation}
  \label{eq:boundary_conditions_gradient_corrector}
  \BLambda(0) = \BLambda(T) = 0.
\end{equation}

Before discussing the gradient, we first calculate the variational derivative of $\hat J$.
As it is customary in the context of optimization problems, 
we express the validity of the GPE~\eqref{eq:state_equation} in the form of a constraint $Z=0$,
with the contraint functional
\begin{align*}
  Z(\blambda,\psi) = i\partial_t\psi+\frac12\Delta\psi-V_\blambda\psi-g|\psi|^2\psi.
\end{align*}
By definition, $\psi_\blambda$ satisfies $Z(\blambda,\psi_\blambda)=0$,
hence
\begin{align}
  \label{eq:danny3}
  \hat J(\blambda) = L(\blambda,\psi_\blambda,p),
\end{align}
where $L$ denotes the 
Lagrangian which was defined in Eq.~\eqref{eq:Lagrange_function} of the main text.
\begin{align*}
  L(\blambda,\psi,p) = J(\blambda,\psi) + \Re\int_0^T\langle p,Z(\blambda,\psi)\rangle\,dt.
\end{align*}
Eq. \eqref{eq:danny3} holds for arbitrary smooth functions $p:(0,T)\to L^2(\R^3;\C)$.
For fixed $p$, differentiation of $\hat J$ in the direction $\delta\blambda$ yields
\begin{equation}
  \label{eq:danny1}
 \begin{aligned}
   & D_\blambda \hat J(\lambda)[\delta\blambda] 
   = D_\blambda L(\blambda,\psi_\blambda,p)[\delta\blambda] + D_\psi L(\blambda,\psi_\blambda,p)[\delta\psi] \\
   & = D_\blambda J(\blambda,\psi_\blambda)[\delta\blambda]
   + \Re\int_0^T\langle p,D_\blambda Z(\blambda,\psi_\blambda)[\delta\blambda]\rangle\,dt \\
   & \quad +D_\psi J(\blambda,\psi_\blambda)[\delta\psi] 
   + \Re\int_0^T\langle p,D_\psi Z(\blambda,\psi_\blambda)[\delta\psi]\rangle\,dt,
 \end{aligned}
\end{equation}
where $\delta\psi$ is the variation in $\psi_\blambda$ induced by the variation $\delta\blambda$ of $\blambda$,
i.e., it satisfies $D_\blambda Z(\blambda,\psi_\blambda)[\delta\blambda] + D_\psi Z(\blambda,\psi_\blambda)[\delta\psi] = 0$
and $\delta\psi(0)=0$.
For simplification of $D_\blambda\hat J$, we choose $p$, which has been arbitrary up to this point,
such that the last two terms in~\eqref{eq:danny1} cancel.
Indeed, taking $p$ as a solution to the terminal value problem~\eqref{eq:adjoint_equation} and~\eqref{eq:terminal_condition_adjoint_equation},
it follows that
\begin{widetext}
\begin{align*}
  & D_\psi J(\blambda,\psi_\blambda)[\delta\psi]
   + \Re\int_0^T\langle p,D_\psi Z(\blambda,\psi_\blambda)[\delta\psi]\rangle\,dt \\
   & = -\Re\left(\langle\psi_d,\psi_\blambda(T)\rangle^*\langle\psi_d,\delta\psi(T)\rangle\right) 
  + \Re\int_0^T\langle p,i\partial_t\delta\psi +\frac12\Delta\delta\psi - V_\blambda\delta\psi - 2g|\psi_\blambda|^2\delta\psi -g\psi_\blambda^2\delta\psi^*\rangle\,dt \\
  &=\Re\int_0^T\langle i\partial_tp+\frac12\Delta p-V_\blambda p-2g|\psi_\blambda|^2p-g\psi^2p^*,\delta\psi\rangle\,dt = 0.
\end{align*}
\end{widetext}
To arrive at this result, we have performed an integration by parts with respect to time,
using the terminal condition~\eqref{eq:terminal_condition_adjoint_equation}
and the fact that $\delta\psi(0)=0$ thanks to the initial condition~\eqref{eq:initial_condition_state_equation}.
In view of these cancellations, equation~\eqref{eq:danny1} simplifies to
\begin{widetext}
\begin{align}
  \label{eq:danny6}
  D_\blambda\hat J(\blambda)[\delta\blambda] 
  = \gamma\int_0^T\partial_t\blambda\cdot\partial_t(\delta\blambda)\,dt 
  - \Re\int_0^T\langle p,\partial_\blambda V_\blambda\cdot(\delta\blambda)\,\psi\rangle\,dt.
\end{align}
\end{widetext}

We are now in the position to calculate the $H^1$-gradient of $\hat J$.
Recall that the Sobolev space $H^1(0,T;\R^m)$ consists of all square integrable functions $\blambda\in L^2(0,T;\R^m)$
that possess a weak derivative $\partial_t\blambda\in L^2(0,T;\R^m)$.
Functions $\blambda\in H^1(0,T;\R^m)$ are actually H\"older continuous, 
and therefore, they have well-defined boundary values at $t=0$ and $t=T$.
It is natural to consider the reduced cost functional $\hat J$ as defined on $H^1_*(0,T;\R^m)$,
which is the affine subspace of functions $\blambda\in H^1(0,T;\R^m)$ 
that satisfy the boundary conditions~\eqref{eq:boundary_conditions_lambda}.
Indeed, any admissible control $\blambda:(0,T)\to\R^m$ must produce a finite value in the penalty term in $J$, 
which implies that $\partial_t\blambda\in L^2(0,T;\R^m)$.
The tangent space to $H^1_*(0,T;\R^m)$, i.e., the space of possible variations $\delta\blambda$,
is the linear subspace $H^1_0(0,T;\R^m)$ of all functions $\BLambda\in H^1(0,T;\R^m)$ with vanishing boundary values, $\BLambda(0)=\BLambda(T)=0$.
This is a Hilbert space with respect to the inner product
\begin{align*}
  (\BLambda_1,\BLambda_2) := \int_0^T \partial_t\BLambda_1(t)\cdot\partial_t\BLambda_2(t)\,dt.
\end{align*}
By definition, the gradient of $\hat J$ with respect to the inner product $(\cdot,\cdot)$ 
is the uniquely determined element $\BLambda\in H^1_0(0,T;\R^m)$ such that
$(\BLambda,\delta\blambda) = D_\blambda\hat J(\blambda)[\delta\blambda]$
for all variations $\delta\blambda\in H^1_0(0,T;\R^m)$.
In view of~\eqref{eq:danny6}, $\BLambda$ satisfies
\begin{widetext}
\begin{align}
  \label{eq:danny7}
  \int_0^T \partial_t(\BLambda-\gamma\blambda)\cdot\partial_t(\delta\blambda)\,dt 
  = -\int_0^T \Re\langle p,\partial_\blambda V_\blambda\,\psi\rangle\cdot(\delta\blambda)\,dt
  \quad\text{for all variations $\delta\blambda\in H^1_0(0,T;\R^m)$},
\end{align}
\end{widetext}
and $\BLambda\in H^1_0(0,T;\R^m)$ induces the boundary conditions~\eqref{eq:boundary_conditions_gradient_corrector}.
To verify that the solution $\BLambda$ to the boundary value problem~\eqref{eq:gradient_corrector} and \eqref{eq:boundary_conditions_gradient_corrector}
satisfies~\eqref{eq:danny7}, 
it sufficies to integrate by parts in the time integral on the left-hand side, 
using that $\delta\blambda(0)=\delta\blambda(T)=0$.

\subsection{Algorithms and implementation}

\subsubsection{Numerical evaluation of the cost functional}
The evaluatation of the reduced cost functional~\eqref{eq:reduced_cost_functional} for a given control curve  $\blambda$
implicitly involves the computation of  $\psi_\blambda$, that is, the solution of the GPE.
No analytical solutions are available in general, so we use a numerical approximation.
For brevity of notation, we write $\psi$ instead of $\psi_\blambda$ in the following.

For the numerical computation of the first term in ~\eqref{eq:cost_functional}, 
that is $1/2 \, \left(1-|\langle\psi_d,\psi(T)\rangle|^2\right)$, 
we have to solve the GPE~\eqref{eq:state_equation} with initial data~\eqref{eq:initial_condition_state_equation} for $t \in [0,T]$. 
Our simulations are performed on the spatial domain 
\[
\Omega = [-L_x/2, L_x/2] \times [-L_y/2, L_y/2] \times [-L_z/2, L_z/2]
\]
with $L_x$, $L_y$ and $L_z$ chosen sufficiently large to capture the significant part of the rapidly decaying solution $\psi$.

For numerical discretization in time, we employ the following time-splitting spectral method \cite{BaJaMa2003}:
\begin{equation}
  \psi(t_{n+1}) \approx e^{- i B_n^+ \triangle t / 2} e^{-i A \triangle t} e^{-i B_n^- \triangle t / 2} \psi(t_n),
\end{equation}
with operators $A = - 1/2 \Delta$, $B^\pm = V_{\blambda^{(n+1/2)}} + g |\psi^\pm_n|^2 $, and with $t_n = n \triangle t$, $n = 0,...,N-1$ s.t.\ $N \triangle t = T$.
Here $\blambda^{(n+1/2)} = 1/2 \, (\blambda(t_n) + \blambda(t_{(n+1)}))$, and the choice of $\psi^\pm_n$ is given below.
Thus, the $n$th time step consists of the following three sub-steps.
First, solve $i \partial_t\psi = (V_{\blambda^{(n+1/2)}} + g |\psi(t_n)|^2)\psi$ for a duration of $\triangle t / 2$ with initial value $\psi(t_n)$;
thus $\psi_n^-=\psi(t_n)$.
The result is used as initial value for the free Schr\"odinger equation $i \partial_t\psi = - 1/2 \Delta \psi$, which is then solved for duration of $\triangle t$;
the result is $\psi_n^+$.
Finally, $i \partial_t \psi = (V_{\blambda^{(n+1/2)}} + g |\psi_n^+|^2) \psi$ is solved with initial value $\psi_n^+$, 
again for a duration of $\triangle t / 2$.
The result of the third sub-step is taken as $\psi(t_{n+1})$. 

The free Schr\"odinger equation is solved using the Fourier spectral method.
To this end, the wave function $\psi$ is interpolated by a trigonometric polynomial 
on the grid points of the cartesian grid
\begin{align*}
(x_{j_x}&, y_{j_y}, z_{j_z}) = \nonumber\\
&(-L_x/2 + j_x \triangle x, -L_y/2 + j_y \triangle y, -L_z/2 + j_z \triangle z),
\end{align*}
where $\triangle x = L_x/J_x$ with $j_x = 0,...,J_x-1$ etc.
Thus, at time $t_n$, the wave function $\psi(t_n)$ is represented 
by a three-dimensional array of complex numbers $\bm{\psi}^{(n)} \in \C^{J_x, J_y, J_z}$.

As \texttt{Matlab} code, the $n$th time step looks as follows:
\begin{equation}
\label{eq:tssm_matlab}
\begin{aligned}
\bm{\psi} &= \exp(-1i * (\bm{V}_{\blambda^{(n+1/2)}} + g * \mathrm{abs}(\bm{\psi}).\string^2) * \triangle t / 2) \,{.*}\, \bm{\psi}; \\
\bm{\psi} &= \operatorname{fftn}(\bm{\psi}); \\
\bm{\psi} &= \bm{M} \,{.*}\, \bm{\psi}; \\
\bm{\psi} &= \operatorname{ifftn}(\bm{\psi}); \\
\bm{\psi} &= \exp(-1i * (\bm{V}_{\blambda^{(n+1/2)}} + g * \mathrm{abs}(\bm{\psi}).\string^2) * \triangle t / 2) \,{.*}\, \bm{\psi};
\end{aligned}
\end{equation}
The array $\bm{M}$ represents the action of the free Schr\"odinger operator.
Due to the vectorized implementation in Matlab this procedure is highly efficient. 


The method is of second order in time and of spectral accuracy in space, provided that
$\psi_0$ and $V_\blambda$ are sufficiently smooth.
In comparison to a finite difference Crank-Nicolson scheme (see for example \cite{BaJaMa2003, HoReBoSc07}),
the solution of a linear evolution equation is avoided, 
and less grid points $J = J_x J_y J_z$ are needed to achieve the same quality of approximation for $\psi$. 

Typically, the numerical costs for our implementation~\eqref{eq:tssm_matlab} are dominated by 
the fast Fourier transforms \texttt{fftn} and \texttt{ifftn}, which are of order $\mathcal{O}(J \log J)$. 
However, in some simulations (Splitting), 
the costs for computing the external potential $V_{\blambda^{(n+1/2)}}$ exceed that of the Fourier transforms.

For the numerical solution of the optimization problem, on the order of 10 to 100 evaluations of the cost functional are needed.
The respective solution of the time-dependent GPE is performed on the graphics processing unit (GPU) of a powerful graphics card.
Thanks to the vectorized implementation~\eqref{eq:tssm_matlab},
it suffices to initialize 
the arrays $\bm{\psi}$ and $\bm{V}$ once at the beginning,
using the \texttt{Matlab} command \texttt{gpuArray}.
For handling the intermediate results or for calling the data in the memory of the main processor at the end of the computation 
we use the command \texttt{gather}. 
The trap potentials need to be updated in each time step.
However, these calculations can be performed in a vectorized way on the GPU as well.

Finally, we compute $1/2 \, \left(1-|\langle\psi_d,\psi(T)\rangle|^2\right)$ 
with $ \psi(T) \approx \bm{\psi}^{(N)}$, using a quadrature formula. 
The integral $\frac\gamma2\int_0^T |\partial_t\blambda(t)|^2\, dt$ is computed by a quadrature formula as well,
using a finite difference formula of second order for the approximation of the time derivative $\partial_t \blambda$.

\subsubsection{Numerical computation of the gradient}

\begin{figure*}[htp]
\centering
	\includegraphics[width=0.8\textwidth]{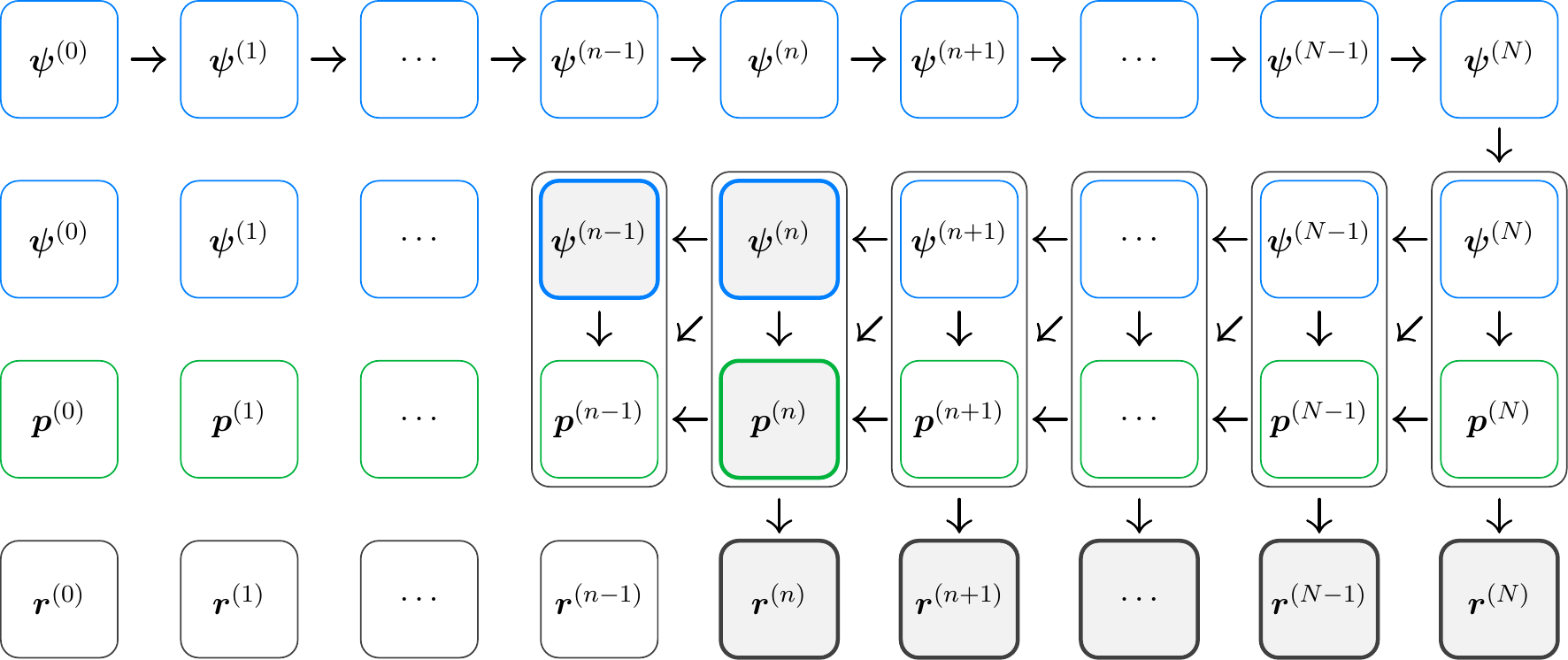}
	\caption{
	Computation of the source term $\bm{r}$ needed to determine the gradient:
	the calculation of the array $\bm{p}^{(n-1)}$ involves only two instances of the wave function 
	$\bm{\psi}^{(n)}$, $\bm{\psi}^{(n-1)}$ and the current adjoint state $\bm{p}^{(n)}$.
	As soon as the approximations of $\bm{\psi}^{(n-1)}$ and $\bm{p}^{(n-1)}$ are available, 
	also $\bm{r}^{(n-1)}$ can be computed.
	At each time step only the grey shaded objects need to be stored in the memory 
	of the graphics card.
	The storage space for $\bm{r}^{(\ell)}$ with $\ell = 0, ..., N$ is negligibly small.
	}
	\label{fig:source_term}
\end{figure*}

According to ~\eqref{eq:gradient_H_0_1},
the $H^1$-gradient $\hat{J}(\blambda)$ is obtained as solution to the second order problem
 \begin{equation}
\label{eq:poisson_equation_r}
  \frac{d^2}{dt^2} \big[\nabla \hat{J}(\blambda)\big] 
  = \bm{r}(\blambda),\quad 
  \bm{r}(\blambda) = \gamma \ddot{\blambda} + \Re \langle \psi, (\partial_\blambda V_\blambda) p  \rangle 
\end{equation}  
subject to the boundary conditions 
$\big[\nabla \hat{J}(\blambda)\big](0) = \bm{0}$ and
$\big[\nabla \hat{J}(\blambda)\big](T) = \bm{0}$.
The time derivatives are discretized by second order finite differences. 

To evaluate the right-hand side $\bm{r}(\blambda)$, 
the functions $\psi$ and $p$ need to be determined for $t \in [0,T]$.
First, the state equation~\eqref{eq:state_equation} is solved as described above.
Then, the adjoint equation~\eqref{eq:adjoint_equation} is solved backwards in time,
for the terminal condition~\eqref{eq:terminal_condition_adjoint_equation}.
For solution of the adjoint equation, a time-splitting method is applied as well:
we alternately solve the equations
$i \partial_t p = V_\blambda p + 2 g |\psi|^2 p + g \psi^2 p^*$ and $i \partial_t p = - 1/2 \Delta p$.
The free Schr\"odinger equation is discretized by the Fourier-spectral method,
and the value of $\partial_\blambda  V_\blambda$ at time $t=(n-1/2) \triangle t$ is computed by means of the complex-step derivative approximation \cite{SquTra1998}.  
 
For integration of the ajoint equation on the time interval $[(n-1)\triangle t,n\triangle t]$, 
an approximation of the wave function $\psi^{(n-1/2)} = 1/2 \, (\psi^{(n-1)} +\psi^{(n)})$ is needed.
Since it is impossible to store the arrays $\bm{\psi}^{(n)}$ for every time step $n=0,...,N$ on the graphics card, 
the state equation is simultaneously solved backwards in time as well.
The procedure is sketched in Fig.~\ref{fig:source_term}:
the calculation of $\bm{p}^{(n-1)}$ involves only two instances of the wave function and the ``old'' adjoint state 
--- that is, $\bm{\psi}^{(n)}$, $\bm{\psi}^{(n-1)}$, and $\bm{p}^{(n)}$.
As soon as the approximations of $\bm{\psi}^{(n-1)}$ and $\bm{p}^{(n-1)}$ are available, also $\bm{r}^{(n-1)}$ can be computed.
In this way it is enough to store at each time step four arrays in three dimensions, 
and the values of all available $\bm{r}^{(n)}$ with $n=0, ..., N$ (the storage space of which is neglegible). 

A further difficulty in the numerical computation of the adjoint equation
arises from the conjugate-complex quantity $p^*$ in $g \psi^2 p^*$.
Without going into details, we refer to the implementation in  \cite{WiBo08}, which can be easily applied to the three dimensional case.
As in the case of the Gross-Pitaevskii equation the computation of $\bm{r}$ can be significantly accelerated by using the graphics card.
Still, the costs for the computation of the gradient are three to four times higher than for the evaluation of the cost functional.

\subsubsection{Computation of the initial and desired final states}

The initial and terminal states $\psi_0$ and $\psi_d$ are assumed to be ground state solutions 
of the stationary Gross-Pitaevskii equation. 
We compute them by imaginary time propagation~\cite{ChiSucTos2000,BaDu2004} (also known as normalized gradient flow): 
the time step  $\triangle t$ in~\eqref{eq:tssm_matlab} is replaced by $-i \triangle t$, 
and the wave function $\phi$ is normalized after every time step.
By using adaptive time stepping, we reach a sufficiently exact solution with justifiable numerical costs.

\subsubsection{Further details of the implementation}

For the numerical solution of the considered optimal control problems we use a personal computer 
(i$7$-$4770$K CPU $@$ $3.50$\,Ghz\,$\times 8$) and Matlab.
The parts with the highest numerical costs, thus the solving of the partial differential equations and the computation of the external potentials, are performed on the graphics card (GeForce GTX TITAN), which accelerates the calculations significantly. 
The evaluation of the Fourier transform, for example, on the finest space discretization can be accelerated by a factor $4$--$6$. 
In this context, it is important to mention that the CPU-version of \texttt{fftn} in Matlab is 
parallelized as well and hence uses all cores available on the CPU.

In general it is useful to initially solve each optimal control problem with a small number of Fourier modes $J_x$, $J_y$, $J_z$ and with a relatively big time step  $\triangle t$. 
Subsequently, the same optimal control problem is solved on a finer mesh grid and with smaller time step, whereby as initial data $\blambda^0$ is used, obtained as approximated solution in the computation before. 
We repeat this procedure until the computed control curve with respect to the old discretization does not differ from the control curve of the finer discretization anymore.

We consider a sequence of discretization parameters
\begin{align*}
&(J_x^{(1)}, J_y^{(1)}, J_z^{(1)}, (\triangle t)^{(1)}) 
\rightarrow 
(J_x^{(2)}, J_y^{(2)}, J_z^{(2)}, (\triangle t)^{(2)}) \\ 
&\rightarrow \ldots
\rightarrow
(J_x^{(M)}, J_y^{(M)}, J_z^{(M)}, (\triangle t)^{(M)})
\end{align*}
with $J_x^{(\ell+1)} > J_x^{(\ell)}$, $J_y^{(\ell+1)} > J_y^{(\ell)}$, $J_z^{(\ell+1)} > J_z^{(\ell)}$ and $(\triangle t)^{(\ell+1)} < (\triangle t)^{(\ell)}$ for $\ell=2,...,M$.
Typically on the order of $10$ to $100$ iterations of the conjugate gradient method are needed for 
solving the optimal control problems on the coarse grid. 
The computational time is of some minutes. 
The numerical costs for the calculations of each single iteration
increase rapidly with each discretization level.
In the same time if one gets near to the local minimum, less iterations for finding the local minimum are required. 
By means of the described strategy each of the presented optimal control problems can be solved in
several hours computing time with respect to the finest discretization level.

\subsection{Numerical solution of the 3D Bogoliubov-de Gennes equations}

For numerical treatment of \eqref{eq:BdG}, we proceed analogously to \cite{RoBoBo2006}:
a change of variables $u = \frac{1}{2} (w_1-w_2)$ and $v = \frac{1}{2} (w_1+w_2)$
transforms the system into:
\begin{equation}
\label{eq:matrix_equation_w1_w2}
-
\begin{bmatrix}
                            & H_0 - \mu + g \phi^2 \\
H_0 - \mu + 3 g \phi^2  & 
\end{bmatrix}
\begin{bmatrix}
w_1 \\
w_2
\end{bmatrix}
= 
\hbar \omega
\begin{bmatrix}
w_1 \\
w_2
\end{bmatrix}.
\end{equation}
A double application of the operator decouples the eigenvalue problem,
\begin{subequations}
\label{eq:eigenvalue_problem_w_1_w_2}
\begin{align}
\label{eq:eigenvalue_problem_w_1}
(H_0 - \mu + g \phi^2) (H_0 - \mu + 3 g \phi^2) w_1 &= \lambda w_1, \\
\label{eq:eigenvalue_problem_w_2}
(H_0 - \mu + 3 g \phi^2) (H_0 - \mu + g \phi^2) w_2 &= \lambda w_2,
\end{align}
\end{subequations}
where $\lambda = \hbar^2 \omega^2$. 
Clearly, it suffices to solve the first eigenvalue problem~\eqref{eq:eigenvalue_problem_w_1}. 


The eigenvalue problem $A w_1 = \lambda w_1$ given in~\eqref{eq:eigenvalue_problem_w_1} 
can only be solved using numerical methods. 
To this end, the operator $A = (H_0 - \mu + g \phi^2) (H_0 - \mu + 3 g \phi^2)$ is discretized via a $6$th-order symmetric finite difference formula. Clearly, $\phi$ und $\mu$ need to be determined in advance and with high precision. 
Here, we solve
\[
H_0 \phi + g |\phi|^2 \phi = \mu \phi, \quad H_0 = -\hbar^2 / 2 m \, \Delta + V
\]
using the same $6$th-order finite difference discretization along with the imaginary time-stepping algorithm (see above).
In this context, the second-order time-splitting method is replaced by the classical Runge-Kutta method of order 4. Subsequently, the chemical potential can be computed using the identity
\[
\mu 
=
\int_{\mathbb{R}^3}
\Big(
\frac{1}{2} |\nabla \phi(\bm{r})|^2 + V(\bm{r}) |\phi(\bm{r})|^2 + g |\phi(\bm{r})|^4
\Big)
\, d\bm{r}.
\]
Once $\phi$ and $\mu$ have been determined we need to solve the discretized eigenvalue problem~\eqref{eq:eigenvalue_problem_w_1}. 

Naturally, we consider the same computational domain
$([-4,4] \times [-15,+15] \times [-2,2]) \, \mu \mathrm{m}^3$
that was used in the splitting experiment in section~\ref{sec:splitting}.
Like in the original experiment we employ $J_x=96$, $J_y=128$ and $J_z=48$
grid points in the respective coordinate directions (in the finest discretization level). 
The resulting large-scale eigenvalue problem is then solved efficiently by means of an iterative algorithm. 
For this purpose we employ the Matlab function \texttt{eigs} which only determines the most relevant eigenvalues and their corresponding eigenfunctions:
the algorithm yields the eigenvalues closest to a specified shift $\sigma$ which we set to a value slightly larger than zero. 
(We are only interested in the first few non-trivial solutions of~\eqref{eq:eigenvalue_problem_w_1} corresponding to the eigenvalues of smallest magnitude.)
The underlying algorithm of \texttt{eigs} requires the repeated solution of the linear system of equations
\begin{equation}
\label{eq:linear_system}
(A - \sigma I) x = b
\end{equation}
for a given right hand side $b$. 
We employ the biconjugate gradients stabilized method (\texttt{bicgstab}) which is implemented in 
Matlab as well. 
Note that $A - \sigma I$ is badly conditioned which is why the \texttt{bicgstab}-routine needs to be 
called with a preconditioner $M = M_1 M_2$, 
i.e. equation~\eqref{eq:linear_system} is effectively replaced by 
$M^{-1} (A-\sigma I) x = M^{-1} b$.
We found that the algorithm converges reasonably fast when the factors $M_1$ and $M_2$ are given by the matrices
$L$ and $U$ obtained from a sparse incomplete $LU$-factorization. 
Such an approximate factorization of $A - \sigma I$ can be computed using another Matlab function called \texttt{ilu}. 
For further information about the Matlab functions mentioned above we refer to the Matlab documentation and the literature cited therein.
The time needed to compute a few eigenvalue-eigenvector solutions of the 
Bogoliubov-de Gennes equations depends strongly on the number of grid points $J_x$, $J_y$ and $J_z$.
For the number of grid points reported above the whole computation takes on the order of five hours computing time utilizing the above mentioned CPU.

\subsection{Extracting the excitation from the time-evolution of the wave-function}

\begin{figure}[htb]
	\centering
		\includegraphics[width=0.45\textwidth]{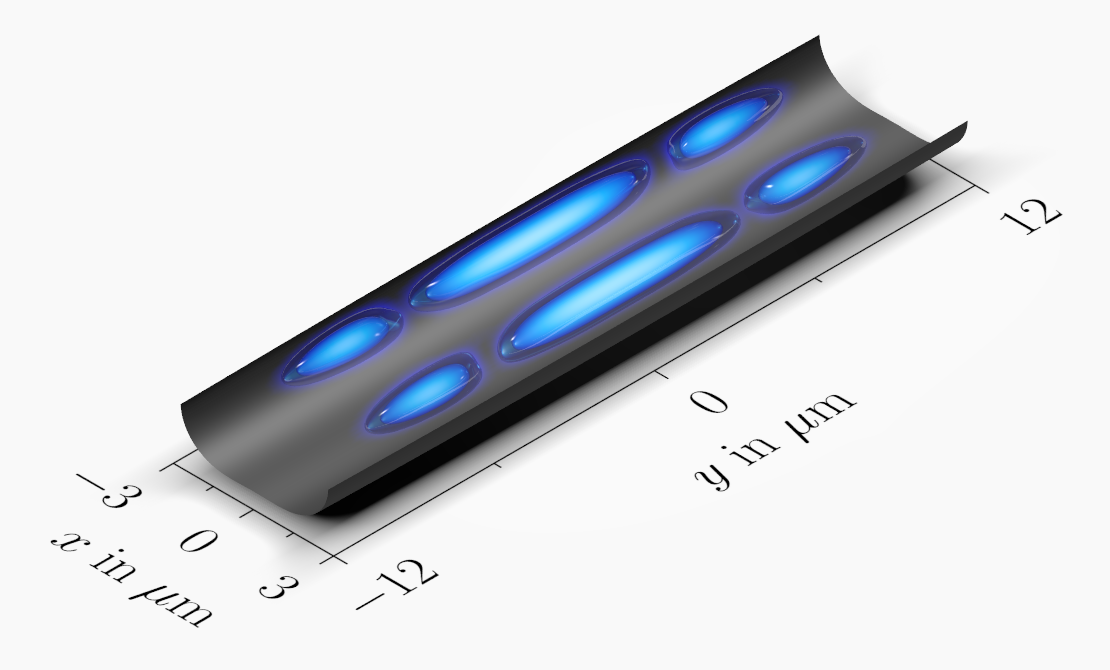}
  		\caption{
		Density of $\triangle \psi(\bm{r},t)$ at $t=9$\,ms.
		}
\label{fig:splitting_Delta_psi}
\end{figure}

The small perturbation which causes the oscillation
of the infidelity 
in the splitting example
can be extracted directly from the time-evolution of the wave function $\psi$.
To this end, we assume that $\psi(\bm{r},t)$ and $\psi_d(\bm{r})$ are almost identical for $t=T$, i.e.,
\[
\psi(t=T) \approx e^{i \theta} \psi_d, \quad \theta = \operatorname*{arg\,min}_{\theta'} \, \| \psi(T) - e^{i \theta'} \psi_d \|.
\]
This assumption is in good agreement with our observations, where the minimum value of the infidelity is 
reached at this point in time.
In analogy to equation~\eqref{eq:perturbation_psi} we define the difference
$\triangle \psi := \psi(\bm{r},t) -  \Phi(\bm{r},t)$, which leads to the result
\[
\triangle \psi(\bm{r},t) := \psi(\bm{r},t) - e^{i \theta} \psi_d(\bm{r}) e^{-i \mu (t-T) / \hbar}, \quad t \geq T.
\] 
Here, we have introduced an additional phase factor $e^{ i \mu T / \hbar}$ in order to take into account that we consider the time-evolution of $\triangle \psi$ starting at $t=T$.
A snapshot of the density $|\triangle \psi(\bm{r},t)|^2$ for $t=9$\,ms is shown in 
Fig.~\ref{fig:splitting_Delta_psi}.
It is quite obvious that the distribution of the density is very similar to the distribution of the density of the second excitation depicted in Fig.~\ref{fig:vsi_solutions_BdG}.

\subsection{One-dimensional approximation for the splitting of a BEC}

\begin{figure*}[b]
	\centering
	\includegraphics[width=0.975\textwidth]{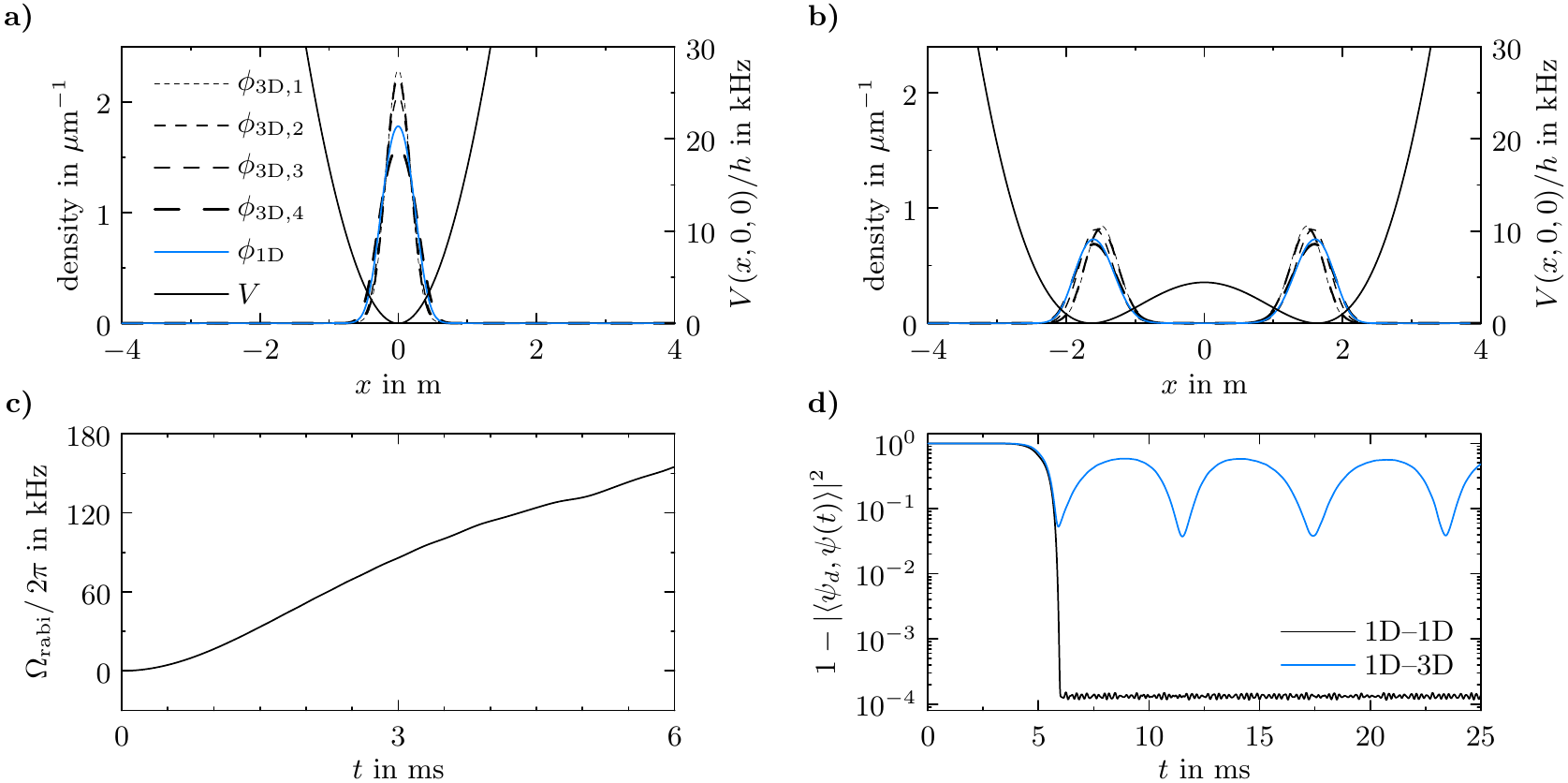}
	\caption{
	Initial state a) and desired state b) of the splitting example along the transversal $x$-direction  
  	The blue solid lines correspond to the eigenstates of the 1D approximation using the 
  	effective coupling constant $g_\mathrm{1d}$.
  	Dashed lines correspond to the eigenstates of the full 3D GPE evaluated along the $x$-direction for 
  	shifted values of $y$ and $z$.
  	Here, 
  	$\phi_{\mathrm{3D},1}(x) = \phi_\mathrm{3D}(x, 7.5\,\mu \mathrm{m}, 1\, \mu \mathrm{m})$,
  	$\phi_{\mathrm{3D},2}(x) = \phi_\mathrm{3D}(x, 7.5\,\mu \mathrm{m}, 0\, \mu \mathrm{m})$,
  	$\phi_{\mathrm{3D},3}(x) = \phi_\mathrm{3D}(x, 0\, \mu \mathrm{m}, 1\, \mu \mathrm{m})$ and
  	$\phi_{\mathrm{3D},4}(x) = \phi_\mathrm{3D}(x, 0\,\mu \mathrm{m}, 0\,\mu \mathrm{m})$.
  	Each wave function has been normalized to unity.	
	c) Optimal control of the Rabi-frequency corresponding to the 1D approximation.
	d) Infidelity (1D--1D) corresponding to the one-dimensional model and infidelity (1D--3D) when the same 
	trajectory of the Rabi-frequency is applied in a simulation using the 3D model.
	}
	\label{fig:splitting_1d_3d}
\end{figure*}

We briefly discuss the 1D approximation for the splitting example. 
In this case, the reduced GPE for the $x$-direction is given by 
\[
i \hbar \partial_t \psi = -\tfrac{\hbar^2}{2 m} \partial_{xx} \psi + V_\lambda(x,0,0) \psi
+ g_\mathrm{1d} |\psi|^2 \psi,
\]
where the effective 1D interaction strength $g_\mathrm{1d}$ is found by integrating out the two transversal dimensions~\cite{Salasnich2002,Olshanii1998}
\begin{equation}
\label{eq:formula_estimate_g_1d}
g_\mathrm{1d} \approx g \int_{-\infty}^{\infty} \int_{-\infty}^{\infty} |\tilde{\phi}(y,z)|^4 \,dy \,dz.
\end{equation}
Here, $\tilde{\phi}(y,z) := \phi(0,y,z)$ corresponds to the normalized ground-state solution of the 
3D model in the $(x\equiv 0)$--plane.

With this approximation we find $g_\mathrm{1d} \approx h \times 1300.44$\,Hz~$\mu$m for $N=2000$ atoms.
This value describes the situation along the whole $x$-axis and also leads to reasonable results away from the center of the cloud, as can be seen from Figs.~\ref{fig:splitting_1d_3d}a-b. 
We then follow the same procedures as in the 3D case to find an optimal control trajectory for the Rabi frequency.

\end{document}